\documentstyle[draft,epsf]{mn}

\def\oii{{\rm O{\sc ii}}}
\def\gsim{\mathrel{\raise0.35ex\hbox{$\scriptstyle >$}\kern-0.6em 
\lower0.40ex\hbox{{$\scriptstyle \sim$}}}}
\def\lsim{\mathrel{\raise0.35ex\hbox{$\scriptstyle <$}\kern-0.6em 
\lower0.40ex\hbox{{$\scriptstyle \sim$}}}}

\title[The History of Star Formation in Rich Clusters]
{Reconstructing the History of Star Formation in Rich Cluster Cores}
\author[Kodama \& Bower]
{
Tadayuki Kodama$^{1,2}$ and Richard G. Bower$^1$\\
$^1$Department of Physics, University of Durham, South Road, Durham DH1 3LE,
UK\\
$^2$Department of Astronomy, University of Tokyo, Hongo, Bunkyo-ku,
Tokyo 113-0033, Japan\\
}

\date{Accepted for publication in Monthly Notices of the Royal Astronomical Society.}

\begin{document}

\label{firstpage}

\maketitle

\begin{abstract}

Our study begins by revisiting the {\it photometric} Butcher-Oemler
effect using data from 7 CNOC clusters at 0.23$\lsim$$z$$<$0.43. We
construct the foreground/background corrected colour-magnitude diagrams
for these clusters. Our analysis shows that the CNOC clusters reproduce
the trend of increasing blue galaxy fraction with redshift as seen by
Butcher \& Oemler. We use these data to investigate the history of star 
formation in clusters by connecting these clusters as snapshots at
different  redshifts. We address two key issues. Firstly, we ask whether
the simple fading and passive evolution of the blue galaxies is
consistent with the properties of galaxies in nearby clusters, such as
the Coma cluster. We find that the evolution of star forming field
galaxies towards redder colour (once the star formation ceases on entry
into the cluster environment) can successfully reconstruct colours and
magnitudes of galaxies in the local cluster. There is no
requirement for widespread disruption of these galaxies. 
Since the blue galaxies fade as they age, the fainter galaxies on the 
present-day colour-magnitude relation tend to have more extended star 
formation histories than their bright counterparts. However, this effect 
is {\it not} sufficient to cause a sizable age
variation for the galaxies along the colour-magnitude relation, implying
that the slope is dominated by variations in metal abundance.

Secondly, we address the nature of the Butcher-Oemler effect itself.
We compare the
distribution of galaxies in the colour-magnitude diagrams and hence infer
the evolution of the rate at which galaxies have arrived in the cluster.
Models in which star formation is abruptly truncated as galaxies are accreted
by the cluster have difficulty in reproducing the observed colour 
distribution. In contrast, if star formation declines on a 1~Gyr timescale
after accretion, the galaxy accretion history we infer is consistent from 
cluster to cluster and matches well the distribution expected in simple
theoretical models.
The Butcher-Oemler effect is thus driven both by the declining star formation
rates of field galaxies and by a decline in the rate at which fresh galaxies
are accreted by the cluster. 

Our study naturally leads to a comparison of the
global star formation histories of galaxies in clusters and the field.
We show that the star formation rate per galaxy mass for galaxies
in cluster cores is significantly smaller than that
of the field environment below $z$$<$1 due to the truncation of star
formation. However, the factor by which star formation is suppressed is
dependent on the cluster accretion history. High quality observations
of clusters at higher redshifts are needed to better define this relation.

\end{abstract}

\begin{keywords}
galaxies: clusters -- galaxies: formation -- galaxies: evolution --
galaxies: stellar content
\end{keywords}

\section{Introduction}

It is 20 years since Butcher \& Oemler (1978; 1984, hereafter BO84)
first reported their startling results on the blue galaxy fractions of
distant clusters.
This and subsequent photometric work suggested that the cores of distant 
clusters at redshift  $z$$>$0.2 contain large numbers of blue galaxies
(Couch \& Newell 1984; Lubin et al. 1996; Rakos \& Schombert 1995),
in stark contrast to the homogeneous red populations of local galaxy
clusters (eg., Bower, Lucey \& Ellis 1992; Terlevich, Bower \& Caldwell 2000).
There is an apparent contradiction between the narrow
colour-magnitude (CM) relation of evolved galaxies in local clusters
(which indicates a long period of passive evolution)
and the blue colours of intermediate redshift galaxies
(which indicates considerable, relatively recent star formation activity).
Bower, Kodama \& Terlevich (1998, BKT98), Smail et al.\ (1998) and 
Balogh et al.\ (2000) have begun the process of 
modelling the combined dynamical and photometric evolution of clusters
in order to reconstruct the galaxies' formation histories. We continue these
themes in this paper.

BO84's original discovery of strong cluster evolution
involved only the photometric properties of galaxies. We will refer to
this aspect as the `photometric Butcher-Oemler effect'.
Dressler \& Gunn (1992) and Couch \& Sharples (1987)
undertook spectroscopy initially with the motivation of confirming 
spectroscopically that the blue galaxies were cluster members.
They discovered that many of the blue cluster galaxies, and some
red ones, had spectra of a type that are rare in local galaxy
samples. These galaxies (referred to as E+A, PSG, a+k/k+a) have 
anomalously strong
Balmer lines indicating an excess of A-stars superimposed on an
K-star type spectrum. The galaxies are produced when star formation
is abruptly truncated (Couch \& Sharples 1987).
The strongest lined examples require a burst of star formation just before
the truncation (eg., Barger et al. 1996; Poggianti \& Barbaro 1996;
Poggianti et al. 1999; Couch et al. 1998).
We refer to the increase in the incidence of such spectra with redshift as the 
`spectroscopic Butcher-Oemler effect'

With the advent of HST, it has become possible to study the morphologies
of galaxies in the distant clusters (Smail et al. 1997; Dressler et al.
1994; 1997; Couch et al. 1994; 1998).
They found that the morphological mix in these
clusters also evolved with redshift, creating a third variant of the 
Butcher-Oemler effect, which we will refer to as the `morphological
BO effect'. In particular, the fraction of S0 galaxies in the cluster cores
decreases rapidly with redshift (Dressler et al. 1997).
Several attempts have been made to connect photometric/spectroscopic
features and the morphology.
Dressler et al. (1994) and Couch et al. (1994) showed that most of the blue
BO84 galaxies are actually normal spirals with the rest being interacting
galaxies or mergers.
Poggianti et al. (1999) compared the spectroscopic
and the morphological effects, suggesting that the time scale of the
morphological transition is longer than the A-type star evolution
since the post star burst galaxies (k+a/a+k) are observed as normal spirals
rather than S0 galaxies.

In this paper, we aim to clearly separate these three different
(but related) evolutionary trends.
We will focus on the {\it photometric} effect, returning
to the questions raised by Butcher \& Oemler's original papers.
The advantage of the photometric approach is that it allows us to study
large, complete samples of galaxies rather than a smaller subset of the 
population that is complicated by target selection and the strength of
spectral features and often limited to brighter galaxies.
Recent studies have tended to concentrate on particular spectroscopic features
to learn details about the nature of the star-burst/truncation phenomena
seen in some cluster members (Barger et al. 1996; Poggianti et al. 1999;
Balogh et al. 1999).
On the other hand, we have made little progress on the evolution
of the cluster population as a {\it whole}, as galaxies accrete from field,
truncate their star formation, and fade and become red towards the CM
relation ridge-line.
A conspicuous contradiction exits between Poggianti et al. (1999) and
Balogh et al. (1999)
over the way star formation is suppressed in the cluster environment.
While Poggianti et al. (1999) require an abrupt end to star formation
activity to explain the incidence of a+k/k+a spectra, Balogh et al. (1999)
favor a slow decline in star formation in order to explain the radial
dependence of the [\oii] emission line strength.

To remedy this situation, we will look at the CM diagrams of the distant
clusters in more detail, examining how the whole cluster population
evolves with time.
{\it The essential idea is that we determine the statistical star
formation histories of galaxies 
in clusters by connecting a series of cluster snapshots at different
redshifts.}
The process has been extensively applied to global galaxy populations
(eg., cosmic star formation history by Madau, Pozzeti \& Dickinson 1998),
but this is the first application specifically targeting cluster galaxies.
We also address how the blue fraction varies with galaxy magnitude and 
with radius from the cluster centre. The variation with
magnitude is of interest given Cowie's claim of cosmic `down-sizing':
the progression of star formation into smaller and smaller units
as the universe ages (Cowie et al. 1996).

Clearly, the history of galaxy accretion will vary from cluster to cluster,
and this will limit the degree to which we can reconstruct cluster galaxy
evolution from a set of observations of different clusters.
However, it is important to distinguish between the growth of the total
cluster mass, which will be dominated by mergers between similar mass units, 
and the accretion of field galaxies. Monte-Carlo simulation of cluster 
growth (Lacey \& Cole, 1993) shows that while the sequence of mergers varies greatly between
clusters, the role of accretion is more stable,
and the fraction of galaxies accreted varies $\sim$ 2 per cent per Gyr.
This is particularly the case if we base our study on X-ray
selected clusters since the existence of a dense intracluster medium
biases against the inclusion of clusters which are not yet dynamically
relaxed (Bower et al. 1997). Nevertheless, differences in cluster
accretion histories can only be eliminated using a much larger
sample of clusters than we have available at present.

It is also important that we try to match the masses of the clusters
in order to form an evolutionary sequence.
The X-ray luminosities of the CNOC clusters used in this study bracket
that of the Coma cluster (\S~2.1). Since the X-ray luminosity function evolves
little over the look-back times involved (eg., Ebelling et al. 1996), the 
clusters should be reasonably well matched in terms of their space density.
Possibly, we would have liked to compare with an even richer local cluster
(\S~2.4). However, although there is no such massive local cluster,
current data on the richness dependence of the CM relation suggests that it
would have even an tighter CM ridge-line (and smaller blue fraction) than 
that of Coma (L\'opez-Cruz 1996). 

Our goals in this paper are two-fold. Initially, we will
make an accurate comparison of the distant and local clusters.
In particular, we will test whether the galaxies in the distant clusters
simply fade and become red after their star formation is truncated or whether
an additional mechanism such as tidal stripping or harassment (Moore et al. 1996) is
required to explain how the distant clusters evolve into clusters
like those at the present-day.  We then go on to explore the
distribution of galaxies in the colour-magnitude plane, and to
address the difference between clusters at different redshifts. The
distribution of galaxies can be used to derive the history of galaxy
accretion from the field and the way the star formation is truncated.
We will show that a consistent picture emerges in which the accretion rate
of cluster galaxies from the field declines smoothly with time. Combined 
with the decline of star formation activity of field galaxies, this naturally
accounts for the Butcher-Oemler effect.

The layout of the paper is as follows.
In \S~2, we present the sources of data used to compose the distant and
local cluster CM diagrams, and outline how we correct
for background and foreground galaxies.
In \S~3, we calculate the blue galaxy fraction from the field corrected
CM diagrams and discuss its dependence on magnitude and the distance from
cluster centre.
In \S~4, we apply our stellar population models to calculate the colour
change (red-wards) and fading of the distant cluster CM diagrams and
compare with a local cluster (Coma).
In \S~5, we investigate the colour distribution of galaxies in more detail,
examining how the galaxy accretion rate must evolve in time.
Our conclusions are summarised in \S~6.

We use H$_0$=50 km s$^{-1}$ Mpc$^{-1}$ and q$_0$=0.1 throughout this paper,
which are the same parameters adopted in the original BO84 work.
The age of the universe is 16.5~Gyr for this parameter set.

\section{Photometric Data for Clusters of Galaxies}

\subsection{Distant Clusters}

\begin{table*}
\caption{
A summary of the cluster sample. Each row shows
(1) the cluster name, (2) redshift, (3) X-ray luminosity in the 0.3--3.5keV
band,
(4) field coverage, (5) look back time at cluster redshift, (6) radial cut
$R_{30}$, (7) BO84 magnitude cut ($M_V$$=$$-$20), (8) slope of the CM sequence,
(9) zero-point of the CM sequence at $r_{\rm cut}$$-$1.5,
(10) BO84 criterion for the blue galaxies defined as the distance from the
CM sequence ($\Delta$($B$$-$$V$)$=$$-$0.2),
and (11) fraction of blue galaxies and Poisson error (1$\sigma$).
}
  \label{tab:clusters}
  \begin{tabular}{lcccccccccc}
  \hline
cluster & $z$ & $L_X$ & $\Delta{\rm RA} \times \Delta{\rm DEC}$ & $t_{\rm LB}$
 & $R_{30}$ & $r_{\rm cut}$ & {\small CM slope} & {\small CM zero-p} & $\Delta$($g$$-$$r$) & $f_B$ \\
 & & 10$^{44}$ erg/s & arcmin$^2$ & Gyr
 & arcmin & mag & & & & \\
  \hline\hline                                                     
A2390       & 0.228 & 23.8 & 46 $\times$ 7     & 3.59 & 2.77    & 20.90
 & $-$0.020 & 0.89 &  $-$0.26 & 0.105 $\pm$ 0.034 \\
MS1008$-$12 & 0.306 &  4.5 & 9.0 $\times$ 7.9  & 4.52 & 2.82$^\dag$ & 21.71
 & $-$0.029 & 1.20 &  $-$0.36 & 0.161 $\pm$ 0.036 \\
MS1224$+$20 & 0.320 &  4.6 & 9.0 $\times$ 7.1  & 4.67 & 2.74$^\dag$ & 21.84
 & $-$0.031 & 1.12 &  $-$0.37 & 0.174 $\pm$ 0.058 \\
MS1358$+$62 & 0.327 & 10.6 & 9.0 $\times$ 23.3 & 4.75 & 3.58    & 21.90
 & $-$0.032 & 1.16 &  $-$0.38 & 0.106 $\pm$ 0.029 \\
MS1512$+$36 & 0.371 &  4.8 & 27.3 $\times$ 8.0 & 5.21 & 3.26    & 22.27
 & $-$0.037 & 1.18 &  $-$0.40 & 0.379 $\pm$ 0.074 \\
MS0302$+$17 & 0.425 &  5.0 & 9.0 $\times$ 7.9  & 5.73 & 2.31$^\dag$ & 22.72
 & $-$0.041 & 1.46 &  $-$0.41 & 0.213 $\pm$ 0.075 \\
MS1621$+$26 & 0.428 &  4.5 & 9.0 $\times$ 23.3 & 5.75 & 2.18    & 22.75
 & $-$0.041 & 1.64 &  $-$0.40 & 0.159 $\pm$ 0.039 \\
  \hline
Coma        & 0.024 &  8.1 & 63 $\times$ 53  & 0.46 & 22.0    & ---
 & ---      & ---  & ---      & 0.056 $\pm$ 0.028 \\
  \hline
  \end{tabular}
Note -- $^\dag$ $R_{30}$ is assumed to 1~Mpc for these clusters due to
the relatively small spatial coverage.\\
\end{table*}

We have chosen to use the CNOC clusters (Yee et al. 1996),
since they form a homogeneous set of intermediate redshift clusters,
selected, with one exception, from the EMSS catalogue of X-ray bright 
clusters (Gioia \& Luppino 1994).
Although the CNOC clusters consist of 16 clusters in total covering
0.17$<$$z$$<$0.55, we have restricted attention to the redshift range
0.23$\lsim$$z$$<$0.43.
We have set the lower limit for the cluster redshift so that the CNOC photometry
data in Gunn $g$ and $r$ (effective wavelengths 4900~\AA\ and
6500~\AA, respectively) bracket the 4000-\AA\ break. Small and well-defined
k-corrections then allow us to reliably compare
the clusters  in rest frame $U$$-$$V$ colours.
We set an upper-limit for the cluster redshift so that the CNOC photometry 
is accurate enough to recover reliable information on galaxy stellar
populations. At $z\sim0.55$, for example,
the photometric error is as high as 0.2 magnitude
even for the relatively bright cluster members ($\sim$$M_*$).
There are 9 clusters within these redshift limits, but the photometric
catalogues are not available for two clusters (MS1231, MS1455) 
at $z\sim0.25$. The final subset of 7 clusters is summarised in
Table~\ref{tab:clusters}.
The clusters are all X-ray bright with 
$L_X = 4-28 \times 10^{44}$ erg s$^{-1}$ (0.3--3.5~keV),
which is comparable to the nearby Coma cluster
($L_X = 8.1 \times 10^{44}$ erg s$^{-1}$)
(Yee et al. 1996; Ebeling et al. 1996; see Table~\ref{tab:clusters} of
this paper).
The advantage of using the CNOC sample is that they obtained $\sim$100--300
redshifts per cluster and also the field of view of both the imaging and
the spectroscopy is relatively large, particularly for four clusters
(Table~\ref{tab:clusters}). Both of these help us in determining cluster 
membership accurately (see Appendix~A).
Furthermore, the wide field enables us to investigate systematic
changes of stellar populations as a function of distance from the cluster
centre (eg., Abraham et al. 1996).

We use the positions, colours and magnitudes in $g$ and $r$ bands,
and spectroscopic redshifts of galaxies, taken from the photometry catalogs
provided by the CNOC group (Yee et al., 1996).
The photometry catalog uses a diameter aperture of $\lsim$6\arcsec,
which corresponds to $\lsim$29--43~kpc for $z$=0.23--0.43.
The completeness limit is about $r$=23.5 magnitude (Yee \& Ellingson 1996),
and we use only those galaxies brighter than this limit in this paper.
Even so, the photometric data has relatively poor accuracy compared to
the data available for local clusters.
The typical photometric error reaches to $\sim$0.2 magnitude at $r$$\sim$23
mag.

\begin{figure*}
\begin{center}
  \leavevmode
  \epsfxsize 1.0\hsize
  \epsffile{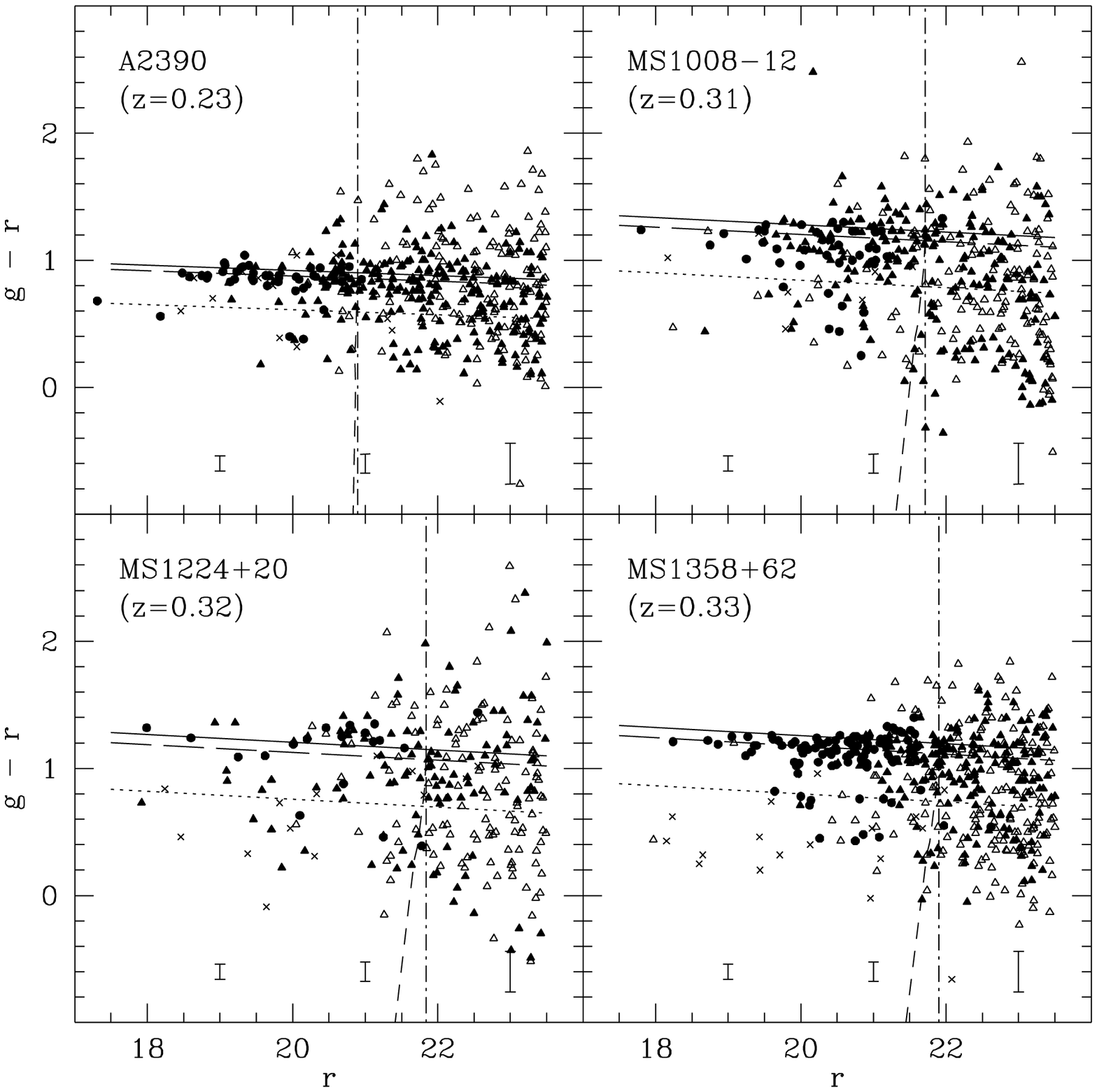}
\end{center}
\caption
{
Colour-magnitude diagrams for A2390 (top left), MS1008$-$12 (top right),
MS1224$+$20 (bottom left), and MS1358$+$62 (bottom right) within $R_{30}$.
The symbols are as follows:-
{\it filled circles}, spectroscopically confirmed members;
{\it crosses}, spectroscopically confirmed non-members;
{\it filled triangles}, statistically plausible members;
{\it open triangles}, statistically subtracted non-members.
The lines in the figure show:- {\it horizontal long dashed line},
the predicted CM red sequence;
{\it solid line}, the red envelope (see text for definition);
{\it dotted line}, colour limit of the blue galaxies corresponding to 
BO84's limit of $\Delta$($B$$-$$V$)=$-$0.2 from the red sequence;
{\it vertical dashed line}, magnitude cut, $r_{\rm cut}$, corresponding to
$M_V$=$-$20 in the rest frame as in BO84. 
{\it Error bars} in the figure illustrate typical $1\sigma$ photometric 
errors and their dependence on magnitude.
}
\label{fig:cm1}
\end{figure*}

\begin{figure*}
\begin{center}
  \leavevmode
  \epsfxsize 1.0\hsize
  \epsffile{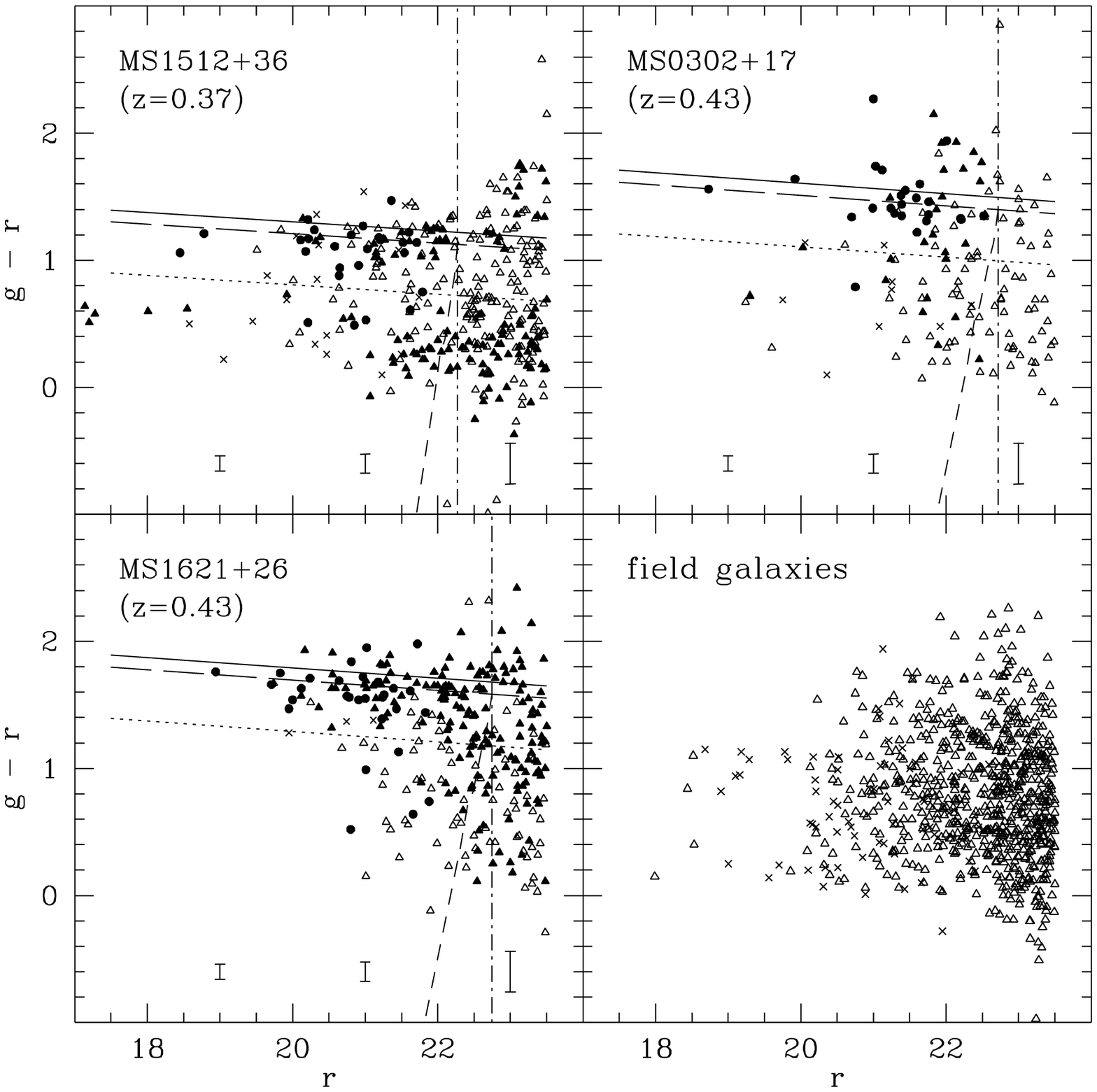}
\end{center}
\caption
{
Colour-magnitude diagrams for
MS1512$+$36 (top left), MS0302$+$17 (top right), and MS1621$+$26 (bottom left)
within $R_{30}$, and for the field taken from the edge of the three 
cluster fields (bottom right; see text). The symbols and lines are explained
in the caption of Fig.~\ref{fig:cm1}.
}
\label{fig:cm2}
\end{figure*}

\subsection{Matching the Cluster Radius}

One of the key ingredients required to make the accurate comparison of
galaxies between local and distant clusters is to match the area
over which the comparison is made.
We use $R_{30}$, the radius from the cluster centre which contains
30 per cent of the whole cluster populations, following BO84.

$R_{30}$ is determined using the galaxy distribution from the cluster
centre after subtracting the average number density of field galaxies
taken from the edge of the cluster field (see below).
Since the shape of the observed field is rectangular, we make up the galaxy
distribution outside of the observed region by
re-sampling the galaxies at the same cluster-centric distance.
We measured $R_{30}$ for four clusters, namely, A2390, MS1358+62, MS1512+36,
and MS1621+26. These are listed in Table~\ref{tab:clusters}, together 
with the concentration index $C$:
\begin{equation}
C=\log (R_{60}/R_{20}),
\end{equation}
 where $R_{20}$ and $R_{60}$ are the radii
which contain 20 and 60 per cent of the whole cluster populations,
respectively.
$C$ is always larger than 0.35, and they are classified as
either compact or intermediate clusters.
Our estimates are in good agreement with those of van der Marel et al.
(1999) except for MS1512+36.
As we will discuss later, MS1512+36 is likely to be contaminated by a
foreground group or cluster, making the $R_{30}$ estimate difficult.
For the remaining three clusters, it is impossible to obtain $R_{30}$
because of their limited spatial coverage.
Therefore we assume $R_{30}$ to be 1~Mpc (H$_0$=50), the typical radius that we measured for the
other clusters. This may be in error by $\sim$30 per cent due to fluctuation
of the background field galaxy density. We will treat these clusters separately in 
our discussion. Note that $R_{30}$ always falls within the original field 
of view.

\subsection{Field Galaxy Subtraction}

The next ingredient that is needed for the accurate comparison between local
and distant clusters is the foreground/background galaxy subtraction.
We use the spectroscopic redshifts as the primary information
cluster membership, where they are available.
We define the cluster members as those whose peculiar velocities are within
$\pm$ 3000~km s$^{-1}$ from the cluster central velocity. Galaxies
outside of this redshift range are assigned to the `field'. 
However, the depth of the spectroscopic redshifts is quite shallow ($r$=21.5),
and there are many galaxies for which spectroscopic redshifts are
unavailable even brighter than this magnitude limit.

We use a statistical method to determine cluster membership
for the galaxies without spectroscopic information.
The details of this method are given in Appendix~A and we give 
a brief summary below.
We take the sample of field galaxies from the edge of the clusters A2390,
MS1358+62 and MS1621+26.
From the distribution of these galaxies on the CM diagram,
we determine the probability that a given galaxy in a cluster with
a given colour and magnitude is a field galaxy.
Using this probability, we apply Monte-Carlo simulations to assign
cluster membership statistically and to subtract
the statistical non-members from the CM diagrams.
The Monte-Carlo simulation is repeated 100 times per cluster and
we average the realisations for the analyses presented in this paper.
Filled symbols in Figs.~\ref{fig:cm1} and \ref{fig:cm2} show the resulting
field corrected CM diagrams for a typical realisation.

\subsection{Local Cluster Comparison}

We use Coma as a representative rich local cluster for comparison
with the CNOC clusters at higher redshifts.
We take the high precision Coma photometric data from Terlevich et al. (2000).
They imaged Coma in $U$ and $V$ bands with a spatial coverage of
3360 arcmin$^2$.
We use a fixed diameter aperture of 25.3~arcsec, both for the magnitudes and
the colours, which corresponds to a physical scale of 17~kpc.
The measurement error is very small, and the typical error in $U$$-$$V$ colour
is less than 0.035 magnitude.
$R_{30}$ for Coma is 22~arcmin or 0.9~Mpc (BO84; Merritt 1987).
We fit the CM relation for galaxies within $R_{30}$ and brighter than
$M_V$$>$$-$20 using the Bi-weight method. We also define a 
`red envelope' at 0.073 magnitude redder than the
fitted CM relation (red sequence).
This colour offset corresponds to a colour difference of the SSP models between
$z_{\rm form}$=2 and 5.4 with solar metallicity.
The CM diagram of Coma within $R_{30}$ is given in the bottom right panel
of Fig.~\ref{fig:cm_faded}.

Note that we will not take into account the dynamical evolution of clusters
in this paper.
An estimate of this effect shows that it is unlikely to be
important if we restrict attention to the region within $R_{30}$.
According to the extended Press-Schechter (P-S) theory (Bower 1991;
Bond et al. 1991; Lacey \& Cole 1993),
a rich cluster comparable to Coma had acquired $\sim$60 per cent of the
present-day mass by $z$=0.33 (q$_0$=0.1).
Judging from the X-ray luminosity (Table~\ref{tab:clusters}),
two of our sample clusters, A2390 and MS1358+62, will evolve into richer 
systems than Coma before the present-day. For these systems, we would have 
liked to compare with an even richer local cluster.
Although there is no such rich counterpart nearby,
data on the dependence of the CM relation with richness suggests that it
would have even an tighter CM ridge-line than that of Coma
(Bower et al. 1992; L\'opez-Cruz 1996).
If this is the case, the recent star formation would be more restricted
and earlier truncation would be required for these clusters
than that discussed in this paper.
In addition, most of the galaxies within $R_{30}$ at intermediate redshifts
will appear in slightly smaller regions than $R_{30}$ at the present-day,
because the cluster will acquire another $\sim$40 per cent of new galaxies
between $z$=0.33 and $z$=0 for example (Bower 1991).
Therefore, the population within $R_{30}$ at a $z$=0.33 cluster should be
actually compared with those within $R_{18}$ at $z$=0, where $R_{18}$
is the radius which contains 18 per cent of the whole cluster population.
However, the colour scatter within $R_{30}$ of Coma is already
so small that it hardly changes within this radius.

\section{The Butcher-Oemler Effect}

\subsection{Blue Galaxy Fraction}

From the field corrected CM diagrams (Figs.~\ref{fig:cm1} and \ref{fig:cm2})
we now calculate the blue galaxy fraction in each cluster.
We follow the original BO84 definition of the blue galaxy fraction:
only galaxies within
$R_{30}$ and brighter than $M_V$=$-$20 in the rest frame
are considered; only those bluer than $\Delta$($B$$-$$V$)=$-$0.2 from the red
sequence are defined as the `blue' cluster members.
We transform these  criteria for blue galaxies into the ones for
the CNOC $g$ and $r$ bands in the observer's frame using the K-corrections
given in Fukugita, Shimasaku \& Ichikawa (1995).
They are based on Kennicutt's (1992a) and Coleman,
Wu \& Weedman's (1980) spectral atlas.
According to Fukugita et al. (1995), $\Delta$($B$$-$$V$)=$-$0.2
roughly corresponds to the colour of Sab-type galaxies in the Hubble
sequence of the present-day galaxies, assuming that the zero-point
corresponds to the E-type colour,
and it can be transformed into $\Delta$($g$$-$$r$) at a given redshift
using the colour difference between E and Sab types.
The BO84 criteria $\Delta$($B$$-$$V$)=$-$0.2 actually falls between
Sab and Sbc types, and we make an interpolation to get the exact
$\Delta$($g$$-$$r$).
This also corresponds to $\Delta$($U$$-$$V$)=0.54 at the present-day,
which is used as a BO84 criterion for our Coma data.
The magnitude cut, $M_V$=$-$20, is also transformed into an apparent
magnitude in $r$-band, $r_{\rm cut}$, for each cluster, assuming the E-type
spectral energy distribution (SED) as in BO84.
The adopted $r_{\rm cut}$ and $\Delta$($g$$-$$r$) for each cluster are
summarised in Table~\ref{tab:clusters}.
These limits are also shown by the dot-dashed and the dotted lines, respectively,
in Figs.~\ref{fig:cm1} and \ref{fig:cm2}.

\begin{figure}
\begin{center}
  \leavevmode
  \epsfxsize 1.0\hsize
  \epsffile{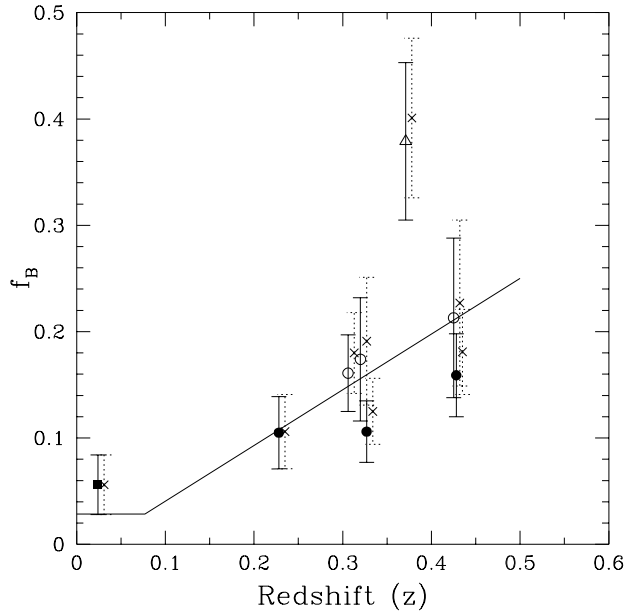}
\end{center}
\caption
{
The blue galaxy fraction ($f_B$) in clusters as a function of redshift
(see text for definition).
The error bars indicate the Poisson errors (1$\sigma$).
The solid line shows the evolution line originally suggested by BO84.
Open circles show the clusters for which we had to assume $R_{30}$
due to the small spatial coverage.
The Coma cluster is plotted as a filled square.
Crosses show the effect of neglecting the colour dependent K-corrections
for the magnitude cut (see text for detail).
MS1512+36 in the open triangle shows an exceptionally high fraction of blue
galaxies, but it is likely to be contaminated by a superposition of a
foreground group/cluster (see text).
}
\label{fig:bo}
\end{figure}

However, the above magnitude cut is not quite correct for the blue galaxies.
A blue galaxy with a given observed $r$ magnitude is actually fainter
in $M_V$ than a red galaxy with the same $r$ magnitude if the $r$-band
in the observed frame corresponds to shorter wavelength than the rest
frame $V$-band, as is actually the case for our CNOC clusters.
Therefore, we apply colour dependent K-corrections using the numbers
for various spectral types given in Fukugita et al. (1995).
The BO84 magnitude cut in $r$-band thus determined is shown by the
slanted dashed line in Figs.~\ref{fig:cm1} and \ref{fig:cm2}.
The deviation from the
constant $r_{\rm cut}$ (dot-dashed) is larger for blue $g$$-$$r$ colours
and increases with redshift.
The simpler definition of the magnitude cut (as is adopted in BO) tends to
overestimate the fraction of blue galaxies by a few per cent
($\Delta$$f_{B}$$\sim$0.2) (see Fig.\ref{fig:bo}).

For each cluster, we define the location of the red sequence on the CM diagram
as a zero-point from which we measure the colour difference.
Since the CM slope is poorly determined with the CNOC photometry,
we use a model prediction, assuming the CM slope is governed by
metallicity variation. We adopted the metallicity sequence model in
Kodama \& Arimoto (1997).
The model slope is calibrated to reproduce the Coma CM slopes in $U$$-$$V$
and $V$$-$$K$ (Bower et al. 1992) at 15~Gyr.
Since the CM slope depends on the aperture within which we integrate
colours, we adopt the 50~kpc aperture models to closely match the CNOC's
large aperture (cf., Kodama et al. 1998).
Using this fixed slope, the CM red sequence is fitted to the observed data
points so that the dispersion around the fit is minimised
after removing the galaxies that are located outside of the one sigma limits
for all the galaxies.
The slopes and the zero-points at $r_{\rm cut}$$-$1.5 ($M_V$$=$$-$21.5)
of the adopted CM red sequences are given in Table~\ref{tab:clusters}.

We also define a `red envelope' which will be used later in \S~4.
We assume that the `red sequence'
corresponds to $z_{\rm form}$=2, supposing it is the typical star formation
epoch of cluster ellipticals
(eg., Bower et al. 1992; Ellis et al. 1997;
Stanford, Eisenhardt \& Dickinson 1998;
BKT98; Kodama et al. 1998; van Dokkum et al. 1998a; 1998b),
and draw the red envelope parallel to the red sequence so that the
envelope is redder than the sequence by the same amount of colour
difference (in $g$$-$$r$) between a SSP model of $z_{\rm form}$=2
($T_G$=12.3~Gyr) and that of $z_{\rm form}$=5.4 ($T_G$=15~Gyr).
The colour difference ranges from 0.05--0.1 magnitude.

Following all these steps, we can now reproduce the definition of the blue 
galaxy fraction ($f_B$) used by BO84, with the addition of a colour 
dependent magnitude cut as discussed above. We determine error bars
by combining the uncertainty in the blue fraction from the statistical
subtraction with a Poisson error based on the numbers of blue galaxies:
\begin{equation}
{\rm error} =  \frac{1}{N_{\rm total}} \sqrt{N_{\rm blue}^{\rm cluster} +
N_{\rm blue}^{\rm field, stat} \cdot
\frac{A_{\rm cluster}}{A_{\rm field}}},
\end{equation}
where $N_{\rm total}$ is the total number of galaxies after field
subtraction, $N_{\rm blue}^{\rm cluster}$ is the number of blue members,
and $N_{\rm blue}^{\rm field, stat}$ is the number of blue field galaxies
that are subtracted statistically, 
$A_{\rm cluster}$ and $A_{\rm field}$ are the area of the cluster 
 and the area of the field sample,
respectively.  The Poisson component always dominates the error.
The blue fraction of each of the clusters is summarised in
Table~\ref{tab:clusters}.

For A2390 and MS1621+26, we can cross-compare $f_B$ with the earlier 
estimates by Abraham et al. (1996) and Morris et al. (1998), respectively.
$f_B$=0.14$\pm$0.05 for A2390 and $f_B$$\sim$0.2 for MS1621+26 at our 
$R_{30}$ are both consistent with our estimate within the errors. 
We can also compare our results for A2390 with the study of a larger 
sample of X-ray luminous ($L_X$(0.1--2.4) $\ge$ 16 $\cdot$ 10$^{44}$
erg s$^{-1}$) clusters at $z$$\sim$0.2 by Smail et al. (1998).
Working wih rest-frame $B$$-$$I$ colours, they found an average blue fraction 
of $0.032\pm 0.069$.
Our estimate for A2390 is somewhat larger than this number, but consistent
at the 1~$\sigma$ level.

Figure~\ref{fig:bo} shows $f_B$ as a function of cluster redshift.
Our results fit the original evolution line (solid line) suggested by BO84
remarkably well.
It is notable, however, that one cluster, MS1512+36 at $z$$=$0.37, has
anomalously high blue galaxy fraction compared to the BO84 line and the
other CNOC clusters.
The CM diagram of this cluster suggests that there may be
a foreground group/cluster superimposed along the line of sight.
As seen in the top left panel of Fig.~\ref{fig:cm2},
there are $\sim$5 bright galaxies ($r$$<$19)
with similar colour ($g$$-$$r$$\sim$0.6) within $R_{30}$
which are not subtracted in the statistical realisations.
This clump appears to continue down to fainter magnitudes and to form
another CM sequence.
Although most of the galaxies on this secondary sequence do not
have spectroscopic redshifts (which makes it difficult to identify the
foreground structure conclusively) there are 5 more galaxies in the entire
cluster field which have similar colours of $g$$-$$r$$\sim$0.6 and 
spectroscopic redshifts in the narrow range 0.163$\pm$0.002.
At this redshift, the colour is consistent with passively
evolving early-type galaxies (Kodama \& Arimoto 1997), which leads
us to suggest that the high value of the blue fraction of MS1512+36
is contaminated by the foreground group/cluster.

\subsection{Dependence on Radius}

\begin{figure}
\begin{center}
  \leavevmode
  \epsfxsize 1.0\hsize
  \epsffile{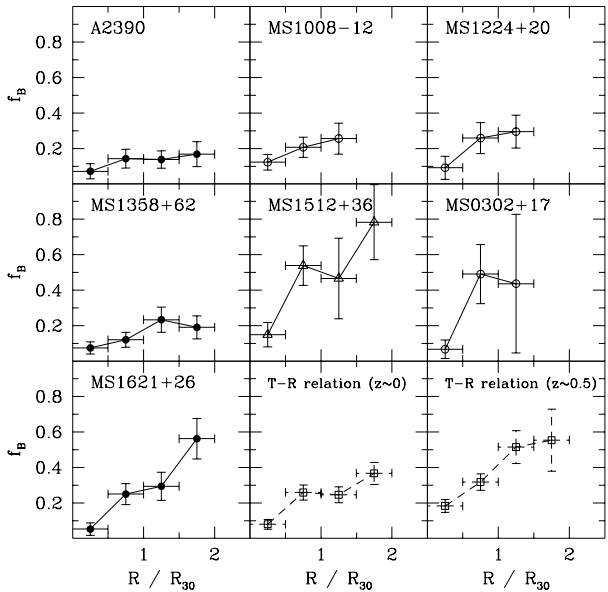}
\end{center}
\caption
{
Dependence of the blue galaxy fraction on the radius ($R$) from the cluster
centre. The radius is normalised using $R_{30}$.
The same symbols are used as in Fig.~\ref{fig:bo} depending on the cluster.
The morphology-radius (T-R) relations at $z$$\sim$0 and $\sim$0.5
(Dressler et al. 1997)
are plotted in the bottom middle and the bottom right panel, respectively,
assuming that the spiral galaxy fraction is directly connected to the blue
fraction and that $R_{30}$=1~Mpc.
}
\label{fig:bo_radius}
\end{figure}

One of the interesting questions concerning the blue galaxies is
their spatial distribution.
If the galaxies are continuously accreted from the surrounding field,
we should expect the blue galaxies to be located in the outskirts of the 
cluster. To test for this radial dependence, we separate the CM diagrams
into 4 bins within 2~$R_{30}$. If the radius exceeds the shorter side of the
rectangular observed strip, we make up the galaxy population outside of the
strip by statistically sampling the galaxies at the same radii within
the observed field.

Fig.~\ref{fig:bo_radius} shows a general trend for $f_B$ to increase
with radius.
This is consistent with the previous results shown by BO84,
Abraham et al. (1996) and Morris et al. (1998).
This trend is to be expected, since it mainly reflects the
morphology-density (or morphology-radius) relation (Dressler et al. 1997)
which shows an increasing (decreasing) fraction of spiral (elliptical)
galaxies with galaxy density or radius. For comparison,
we reproduce the morphology-radius (T-R) relation by Dressler (1997)
for the local clusters ($z$$\sim$0) and for the intermediate redshift
clusters ($z$$\sim$0.5) in the bottom middle and the bottom right panels
of Fig.~\ref{fig:bo_radius}, respectively. Our sample clusters fall between
these two redshifts.
Here we assume the blue fraction is equivalent to the
spiral fraction, since the BO84 colour cut for the blue galaxies
corresponds to Sab type in the local universe as already mentioned earlier
(\S~3.1; Fukugita et al. 1995). 
The slope of these T-R relations are steep enough to explain the radial
dependence of the BO blue fraction.
This radial dependence also illustrates how important it is that distant
and local clusters are compared within equivalent radii.

\subsection{Dependence on Magnitude}

\begin{figure}
\begin{center}
  \leavevmode
  \epsfxsize 1.0\hsize
  \epsffile{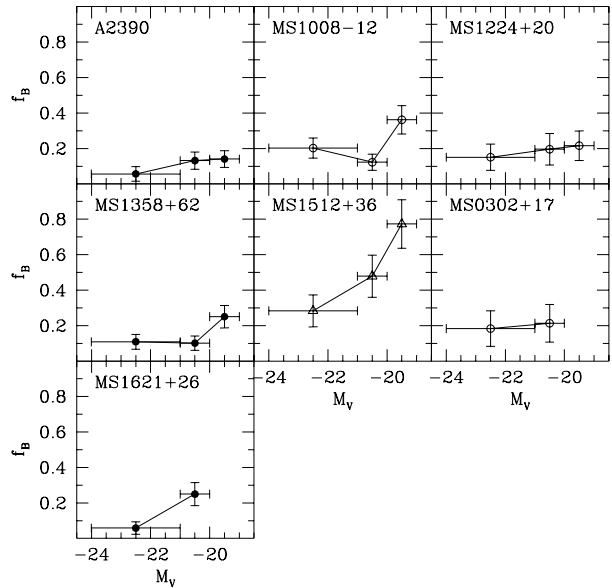}
\end{center}
\caption
{
The dependence of the blue galaxy fraction on absolute magnitude. 
The fiducial magnitude cut for $f_b$ corresponds to $M_V$=$-$20.
The same symbols are described in Fig.~\ref{fig:bo}.
The points at the faintest magnitude bins for the two highest redshift
clusters are dominated by the large photometric errors ($\sim$0.2) and are
not shown.
}
\label{fig:bo_maglim}
\end{figure}

Another important issue is the magnitude dependence of the blue fraction
as shown in Fig.~\ref{fig:bo_maglim}. In most of the clusters, there is 
only a weak trend for the blue galaxy fraction to increase towards 
fainter magnitudes.
MS1512+36, which has the strongest trend, is likely to be contaminated
by a foreground group/cluster (\S~3.1).
The overall impression is that the magnitude dependence is weak
especially for the bright magnitudes ($M_V$$>$$-$20).
BO84 found no systematic change in blue fraction over a similar magnitude
range, except for their brightest magnitude bin ($-$23$<$$M_V$$<$$-$22.5),
where the blue fraction is smaller. However, these results do not mean that 
there is little {\it mass} dependence of the star formation histories
(Smail et al., 1998, Poggianti et al., 1999). 
In order to discuss the intrinsic mass dependence we need to consider the 
fading of the blue galaxies after they enter the cluster, since the 
star forming galaxies are brighter in $M_V$ than the red galaxies of the 
same mass.
We will come back to this point later in \S~4.3 by showing the
properties of the galaxies as a function of the `present-day' magnitude.
This approach will give us a direct view to the mass dependence of star
formation history.

\section{From intermediate redshift clusters to the Coma cluster}

In this section, we address the fate of the blue galaxies seen in the
intermediate redshift clusters. In particular, we investigate whether 
they become incorporated into the tight CM relation seen in the
present-day local clusters.

BKT98 discussed this issue, introducing the models where the star formation
is truncated at the time of galaxy accretion from the surrounding field
to clusters. By estimating the distribution of the truncation epoch by the 
extended P-S theory, they explored the allowed range of star formation 
histories in rich clusters.
In this paper, we discuss this problem in further detail, taking into
account the fading of the galaxies and their colour
evolution after star formation ceases. Our approach extends the 
analysis of Smail et al. (1998) who addressed the evolution of galaxies
in a sample of X-ray luminous clusters at $z\sim 0.2$.
Instead of using the predicted galaxy accretion rate at the first place,
we start from the real galaxy distribution on the CM diagrams of
intermediate redshift
clusters, apply the star formation truncation model for the individual
galaxies, and sketch the whole movement of cluster galaxies on the
CM diagrams down to the present-day. Our approach does {\it not} assume
a particular galaxy accretion history for the clusters:
we discuss the limits that can be placed on the accretion history in \S5.

In the following sections, we concentrate on the results for the three 
clusters for which the background subtraction is most reliable: 
A2390, MS1358+62, and MS1621+26, (see \S~2.3 and Appendix~A). MS1512+36
has been omitted because of the probable contamination by a foreground
system. The three other clusters from the CNOC sample have only a
relatively small area of imaging making the background subtraction 
unreliable. Nevertheless the results for these clusters agree qualitatively
with the systems on which we now concentrate.

\subsection{Modelling Galaxy Evolution Vectors}

\begin{figure}
\begin{center}
  \leavevmode
  \epsfxsize 1.0\hsize
  \epsffile{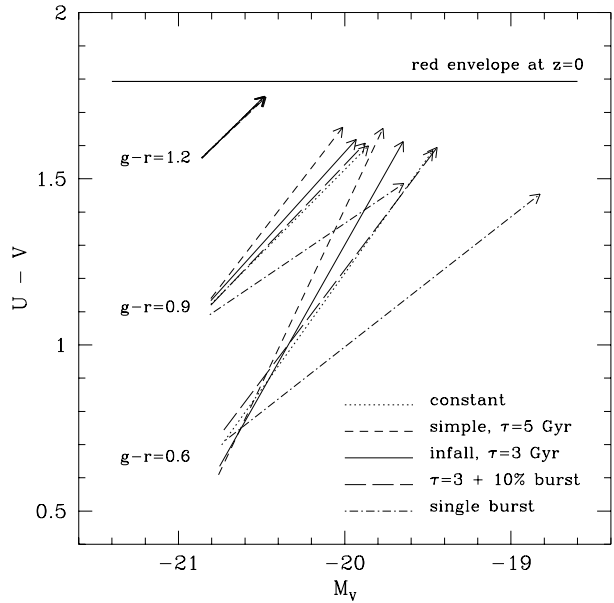}
\end{center}
\caption
{
Evolution vectors, showing the fading and colour evolution 
after star formation ceases, in the rest frame $U$$-$$V$ vs.\ $M_V$ diagram.
The length of the vector shows the evolution expected between $z$=0.33 
and $z$=0. 
Three sets of vectors are plotted to show the evolution of three galaxies 
with different colours at $z$=0.33; ($r,g-r$)=(21,0.6),(21,0.9), and $(21,1.2)$.
For each galaxy five vectors are shown, correspond to different star formation 
histories
prior to the truncation event:-
(1) {\it dotted line}, constant star formation;
(2) {\it short dashed line}, exponentially decaying star formation
with an e-folding time of $\tau$=5~Gyr (simple model);
(3) {\it solid line}: gas infall model with $\tau$=3~Gyr
(see Appendix~B for details);
(4) {\it long dashed line},  as (3), but with a 10 per cent star burst
added just before truncation;
(5) {\it dash-dotted line}, single star burst just before the truncation.
We assume that the star formation starts at $t_{\rm LB}$=15~Gyr ($z$=5.4)
for all the models except (5) and terminates abruptly.
If star formation decays slowly after the truncation event,
the direction and the length of the vectors is hardly changed, although
the speed of evolution along the vector immediately after truncation
does change significantly.
The horizontal solid line shows the red envelope at $z$=0.
}
\label{fig:vector}
\end{figure}

Initially, a difficulty with this approach {\it appears} to be that the 
photometric evolution of a galaxy after the end of its star formation,
will depend on its star formation history prior to the event.
Fortunately, this is not the case: so long as a fraction of stars are old,
the passive evolution phase is only weakly dependent on the previous
history. Fig.~\ref{fig:vector}
shows some examples of evolution vectors from $z$=0.33 to 0
for three galaxies with different colours ($g$$-$$r$=0.6, 0.9, and 1.2).
For each galaxy, we plot 5 different vectors which correspond to
different star formation histories. Star formation is supressed prior 
to $z$=0.33 for all the models, with the truncation time ($t_{\rm trunc}$) 
being chosen so that each model has the same $g$$-$$r$ colour at the moment
of observation. 
Apart from the single burst model which gives the maximum fading and
the bluest colour at the present-day, all the other models show similar
evolution vectors on the CM diagram in spite of large variations
in their previous star formation histories. This remains true even if a 
burst of star formation is included just before the truncation epoch.
As a result, the evolution of the cluster members on the CM diagrams 
depends little on their prior star formation history. 

We use the gas infall model (described in detail in Appendix~B) to
parameterise the star formation of a galaxy before it enters the cluster
environment. In outline, gas inflow onto the galaxy determines its star
formation timescale and hence the distribution of field galaxy
colours. The distribution of gas infall timescales is set so as to match 
the observed colours of the field galaxies in the CNOC sample
and the global star formation in the Universe (Madau et al. 1998).
Once a galaxy is accreted by the cluster, star formation is truncated, 
possibly after a short burst (as described in Appendix~C). 
Adding the burst helps to explain the frequent appearance
of strong Balmer absorption features in the integrated spectra of galaxies
in the intermediate redshift clusters (Barger et al. 1994; Poggianti et al.
1999; BKT98; but see also Balogh et al. 1999).
Although there is ambiguity in the star formation and the gas infall time
scales and the burst strength here,
the subsequent evolution is rather insensitive to these values unless
the burst population dominates.
We assign the star formation/gas infall time scale $\tau$ and the truncation
time $t_{\rm trunc}$ for each galaxy to reproduce its colour in $g$$-$$r$
at the cluster redshift as explained in Appendix~D in detail.
Our approach differs from that of Smail et al. (1998) who applied a
similar kind of truncation model, under the assumption that star
formation was terminated in {\it all} the cluster members at the same
epoch. Instead, we allow $t_{\rm trunc}$ to vary from galaxy to galaxy 
reflecting the way the cluster is being built up continuously over time.
With this parameterisation,
we can later discuss the galaxy accretion history which forms the clusters
(\S~5).
Once the $\tau$ and $t_{\rm trunc}$ is assigned for each cluster member,
the model gives its colour and magnitude evolution in rest frame $U$$-$$V$
colours and absolute $M_V$ magnitudes.

The average metallicity of stellar populations varies with magnitude
along the CM sequence (Kodama et al. 1998; Terlevich et al. 1999;
Kuntschner 2000). This variation can either be incorporated into the
model fading vectors, or the observed galaxy colours can be corrected to a
fiducial metal abundance (as in BKT98). Since the fading vectors depend 
little on metallicity around the solar value ($Z$=0.02), it is simpler to
adopt the second approach.
Therefore, we corrected the observed colours for the CM slope as follows:
\begin{equation}
(g-r)_{\rm corrected}=(g-r)_{\rm observed}-{\rm slope}\times(r-r_0)
\end{equation}
where $r_0$=$r_{\rm cut}$$-$1.5 which corresponds to $M_V$=$-$21.5
in the rest frame.
Note that although we can correct the systematic trend as a function of 
luminosity, we neglect the effect of the
age-metallicity conspiracy (Worthey, Trager \& Faber 1996; Trager 1997;
Shioya \& Bekki 1998) in which
young galaxies are supposed to be more metal rich (and hence the age
and metallicity effects on galaxy colours may cancel out each
other to some extent).
Even if this is the case, this effect would be important only for the red
galaxies on the CM sequence. We will return to consider the implications of 
the CNOC clusters for age variations along the CM relation in \S~4.3.

Before applying the truncation model, we have to correct a small zero-point
mismatch between the observed colours and the model by comparing between
the colour of the red sequence at $r_0$ and the SSP model with solar
metallicity and $z_{\rm form}$=2.
We shifted the observed $g$$-$$r$ colours (already corrected for the CM slope)
by 0.0--0.15 magnitude.
Although this is probably due to the zero-point uncertainties in
the model and the observation, it may also be due to the metallicity
mismatch since we fixed the model metallicity at solar metallicity.

A further problem is that there are a few galaxies which have been
statistically assigned cluster membership, even though they are redder than the
CM sequence. These objects cause problems because they are difficult to
model within our stellar population code. Some of these galaxies may be 
background objects which have not been removed by the statistical
subtraction process. Alternatively, these red objects
are likely to be scattered from the red sequence due to the relatively
large photometric errors ($\sim$0.2~mag) toward fainter magnitudes.
Unlike galaxies scattered blueward of the sequence, the offset of these
galaxies would remain large as we evolve the cluster forward in time,
yet there are no such galaxies seen in the Coma cluster.
We have therefore adopted the scheme of initially correcting these red
galaxies back onto the CM relation sequence. A similar correction is not
required for the blue galaxies since the differences due to photometric
errors decline as the stellar population ages.
It should be noted that the scatter of the red objects from the CM sequence
is not large enough to affect the blue galaxy fraction ($f_B$) significantly
--- as can be seen by comparing the deviation of red and blue objects from
the CM sequence in Figs.~\ref{fig:cm1} and \ref{fig:cm2}. 
Most of the red objects fall within $\Delta$($g$$-$$r$)$<$0.4 from the CM
red envelope, of which most of the spectroscopically confirmed members are
within $\Delta$($g$$-$$r$)$<$0.2.
It is possible these red galaxies are dusty spirals.
If these have extensive on-going star formation, we will tend to 
underestimate the fraction of star forming galaxies in the cluster.
We discuss the consequences of dusty objects within clusters in \S~6.

\subsection{Evolving the Distant Clusters Forward in Time}

\begin{figure*}
\begin{center}
  \leavevmode
  \epsfxsize 1.0\hsize
  \epsffile{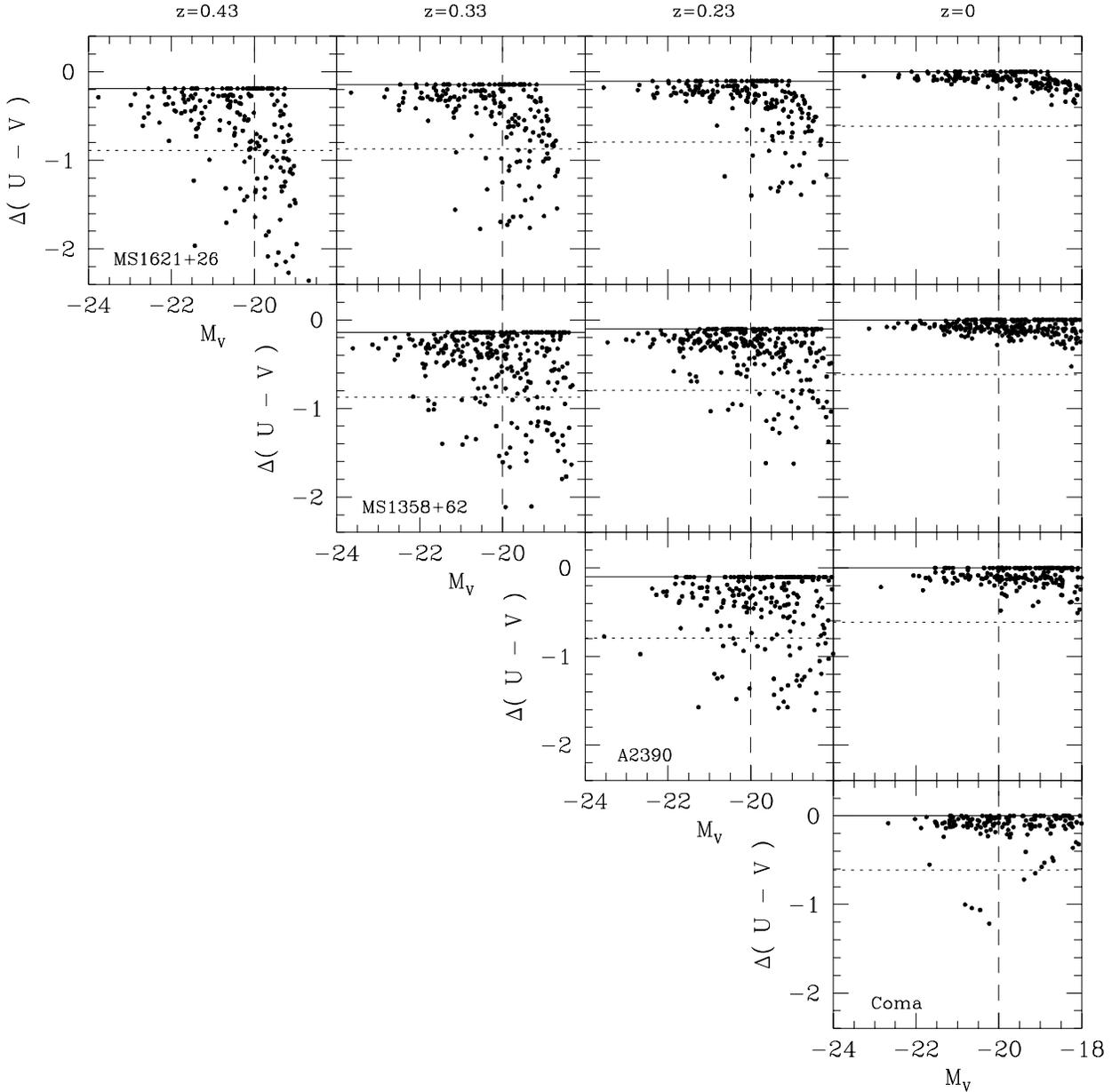}
\end{center}
\caption
{
Evolution of the cluster galaxies within $R_{30}$ forward in time
on the CM diagrams.
From the left panel to the right, redshift is decreasing as 0.43, 0.33,
0.23 and 0 (an evolutionary sequence).
Star formation is truncated after the accretion of galaxies from the field.
The truncation epoch ($t_{\rm trunc}$) is assigned for each galaxy to reproduce
its $g$$-$$r$ colour at cluster redshift.
This is an example of the Monte-Carlo simulations using $e_{\rm loss}$=0 and
$e_{\rm burst}$=0.01 (Model~(b)).
As we propagate the cluster forward in time, no further galaxy accretion 
occurs. The effect of accretion can be seen by comparing the panels
vertically (ie., at the same redshift).
The Coma cluster is plotted in the bottom right panel for comparison with the
$z$=0 models of the CNOC clusters.
Colours are presented with respect to the red envelope (solid line)
to eliminate the metallicity effect. The dotted line and the dashed line
represent the BO84 criterion of the blue galaxies which corresponds to
$\Delta$($B$$-$$V$)=$-$0.2 and $M_V$=$-$20, respectively.
}
\label{fig:cm_faded}
\end{figure*}

We are now in a position to simulate the photometric evolution of the 
cluster galaxy population.
Fig.~\ref{fig:cm_faded} illustrates the evolution of the three clusters
forward in time down to the present-day (from the left panel to the right).
All the cluster members within $R_{30}$ are plotted after the field
subtraction.

Figure~\ref{fig:cm_faded} is an example of the Monte-Carlo simulations
using $e_{\rm loss}$=0 and $e_{\rm burst}$=0.01 (Model (b) in Appendix~C).
The colour is presented with respect to the red envelope at $z$=0 as
discussed in \S4.1. Note that as we evolve the cluster forward in time
(the panels to the right of the initial data), there is no further accretion.
Within a vertical row, the differences between the panels reflect the 
accretion that has occured in the younger cluster.

If we compare the initial CM diagrams (left end panels) of those
intermediate redshift clusters with that of the Coma cluster
(bottom right panel) in the same coordinates, it is obvious that there
are more blue galaxies in the distant clusters and the scatter
around the CM red sequence is larger.
However, if we trace the movement of those blue galaxies forward in time,
we see that the galaxies become redder rapidly as time passes and
all get redder than the BO84 criteria of blue galaxies (dotted line) in only
a few Gyr at most after the cluster redshifts.
As a result, the colour scatter of the CM relation becomes small
enough by the present-day, especially at the bright end ($M_V$$<$$-$20),
to be consistent with the observed scatter in Coma. A more precise
comparison is not justified by the photometric accuracy
of the CNOC data and our consequent treatment of objects scattered
redward of the initial CM relation.
If galaxy harrassment accompanies the galaxy accretion, stripping
off some stars from the galaxy (as suggested by numerical simulations:
Moore et al. 1996; Moore et al. 1999a), the blue galaxies could fade even
further, resulting in a smaller present-day colour scatter at the bright
end of the CM sequence.
On the basis of this comparison, such an effect is not required by the data,
however.

\subsection{The Mass Dependence of Star Formation Histories}

\begin{figure}
\begin{center}
  \leavevmode
  \epsfxsize 1.0\hsize
  \epsffile{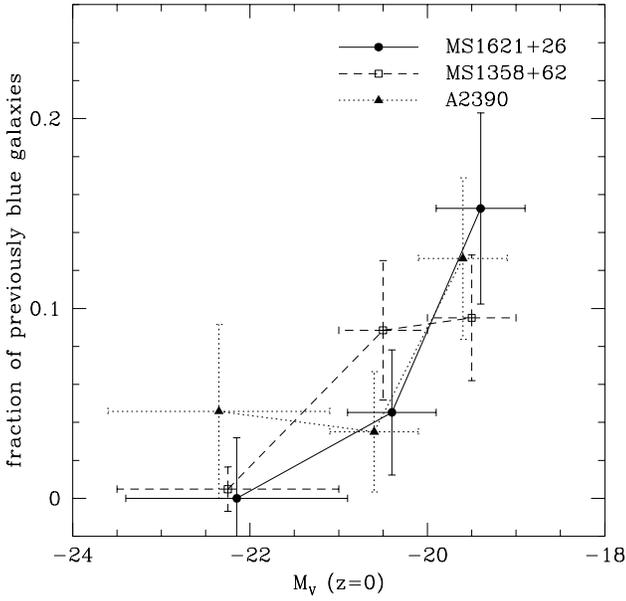}
\end{center}
\caption
{
Fraction of galaxies that are identified as blue galaxies
at the observed epoch plotted as a function of the present-day ($z=0$)
magnitude. This realisation results from adopting Model~(b)
as described in Appendix~C.
The points are slightly offset horizontally to avoid their overlapping,
ie., $-$0.1 for A2390 and $+$0.1 for MS1621+26.
}
\label{fig:downsizing}
\end{figure}

Not only do the galaxies become redder, but they also get fainter after the
truncation event. In fact, most of the blue galaxies which pass the
BO84 criterion at the cluster redshifts fade rapidly and evolve to
red faint galaxies with $M_V$$>$$-$20 ($M_*$+1) by the present-day
as shown in Fig.~\ref{fig:downsizing}
(Fig.~\ref{fig:cm_faded} shows the trajectory of galaxies on the CM diagrams).
The fraction of blue galaxies is now plotted as a function of the present-day
magnitude instead of the magnitude at the observed epoch (as in
Fig.~\ref{fig:bo_maglim}).
Therefore, the blue galaxies at intermediate redshifts
are {\it not} the counter-parts of
the bright early-type galaxies seen in present-day clusters including S0's
(Smail et al., 1998).
This is consistent with Kelson et al.'s (1999) and Poggianti et al.'s (1999)
result derived from the spectroscopic signatures of star formation activity.
It would appear that bright (hence massive) early-type galaxies should have
completed their star formation well beyond $z$=0.4, whereas fainter
(hence less massive) galaxies have had more extended star formation.
This seems to imply that the cosmic 
down-sizing hypothesis (Cowie et al., 1996; Cowie, Songaila \& Barger 1999),
where star formation propagates to smaller units as the Universe ages,
applies to cluster galaxies as well as galaxies in the field.
By extending this analysis to include higher redshift clusters, it
should become possible to more strongly
constrain the formation of massive cluster members.

\begin{figure}
\begin{center}
  \leavevmode
  \epsfxsize 1.0\hsize
  \epsffile{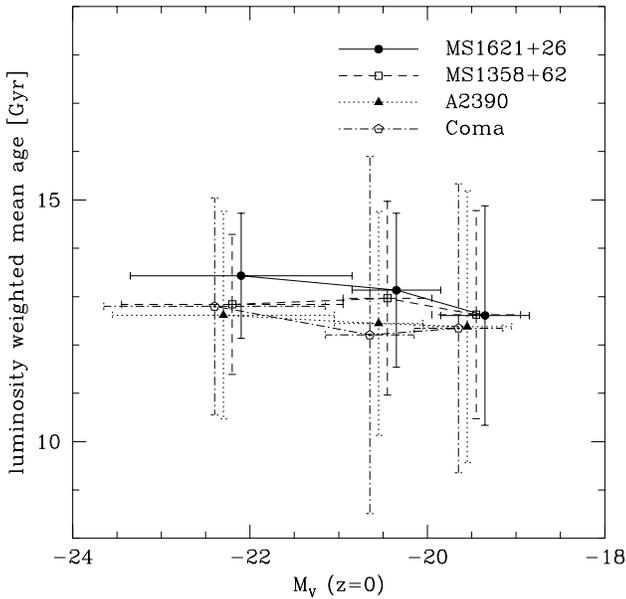}
\end{center}
\caption
{
Luminosity weighted mean age at the present-day as a function of
the present-day magnitude.
Model~(b) is used (see \S~5.1).
The points are slightly offset horizontally to avoid their overlapping,
ie., $-$0.15 for Coma, $-$0.05 for A2390, $+$0.05 for MS1358+62, and
$+$0.15 for MS1621+26.
Error bars show the scatter (1~$\sigma$) within each magnitude bin.
}
\label{fig:tlwm}
\end{figure}

Since the fraction of recently accreted galaxies increases at faint
magnitudes, the resulting age bias may contribute to 
the CM slope. However, if the luminosity ($B$-band) weighted mean age
of the galaxies ($t_{\rm lwm}$) measured at the present-day is plotted
as a function of galaxy luminosity, the systematic trend is very weak;
only $\sim$ 1--2~Gyr difference at most over 3--4 magnitude range of the
present-day CM relation (Fig.~\ref{fig:tlwm}).
The lack of magnitude dependence in luminosity weighted mean age reflects
the fact that $t_{\rm lwm}$ gets old very rapidly once the star formation
is truncated and is rather insensitive to a small amount of 
recent star formation (BKT98).
However, the systematic age difference is sensitive to the CM slope
that we have subtracted to correct the metallicity effect.
To see the maximum age variation that can be allowed along the CM relation,
we repeated the simulation without correcting the CM slope.
All the systematic effects are now attributed to age.
Even in this extreme case, the age difference along the present-day
CM relation is only $\lsim$ 3~Gyr, or less than $\sim$20 per cent, and
fails to reproduce the observed CM slope in Coma: at most only
a quarter of the aperture corrected CM slope can be accounted for in this way.

\section{Implications for Star Formation Histories in Cluster Galaxies}

In the previous section, we have modelled the photometric evolution of 
the clusters' galaxy populations in order to evolve the system forward
in time. In this section, we look at the process from the other view-point: 
using the observed colour distribution to infer star formation and
truncation histories in the cluster environment from high redshifts
down to the observed epoch. 

\subsection{Assigning Star Formation Histories}

We proceed as follows. As a starting point, we adopt a mix of galaxy
star formation histories that are indicated by observations of distant
field galaxies: the details are described in Appendix D. For each
cluster, we are then able to infer a distribution of truncation
times (following the scheme outlined in Appendix~D) from 
the observed colour distribution. The whole process is treated in
a Monte-Carlo sense, and 100 realisations are combined to produce the 
final distribution of truncation times. For example, a galaxy with only
a small blueward deviation from the red CM relation sequence might either be
the result of a red (low star formation rate) field galaxy that arrived 
in the cluster recently, or of a blue (high star formation rate) field
galaxy that arrived long before the observed epoch.  Throughout, we will
assume that the truncation of star formation is associated with the 
build up of the cluster from accreting field galaxies.
Whereas we found that the direction of photometric evolution on the
CM diagrams in the passive evolution phase has little model dependence,
the time scale of the evolution immediately after the truncation epoch
(and hence the colour distribution) does have considerable dependence on the
way the star formation is truncated (for example the burst strength and
the level of residual star formation). It is therefore necessary to consider
different truncation schemes and to compare the cluster formation
histories of the different clusters.

We present results based on two particular models. 
In Model~(a), we consider a violent event that might be expected to
result from ram-pressure stripping
(Abadi et al. 1999; Fujita \& Nagashima 1999)
or harassment (Moore et al. 1996) of the galaxy as it arrives in the 
cluster. We assume that half of the gas in the disk is consumed in a 
strong burst of star formation and that the remaining gas is
lost from the system and no residual star formation is present afterwards.
This model is intended to reproduce the burst models used to explain the
spectroscopic properties of E+A galaxies (eg., Couch \& Sharples 1987;
Poggianti et al. 1999).
For the second model, Model (b), we assume that there is no gas lost
from the disk and that only 1 per cent of disk gas is converted into 
stars by a burst.  In this case the dominant effect of the cluster
is to suppress star formation by `suffocation' --- removing the halo
of material that would otherwise have been slowly accreted by the
galaxy (Larson, Tinsley \& Caldwell 1980; Balogh et al. 2000).
All the remaining gas is consumed exponentially with an e-folding time of 
1~Gyr (half the average star formation timescale for field galaxies).

The two models we have chosen are intended to cover the range of
possible scenarios. In particular, it should be noted that if the burst
strength is made weaker in Model (a), the model struggles to reproduce
the population of blue galaxies; if the star formation decay timescale
is made longer in Model (b), it becomes difficult to reproduce a
sufficient population of red galaxies in the cluster. Nevertheless
models that lie between the two possibilities selected can successfully
reproduce the observed CM diagrams with accretion histories that are also 
intermediate between the two models.

\subsection{The History of Galaxy Accretion}

\begin{figure*}
\begin{center}
  \leavevmode
  \epsfxsize 0.48\hsize
  \epsffile{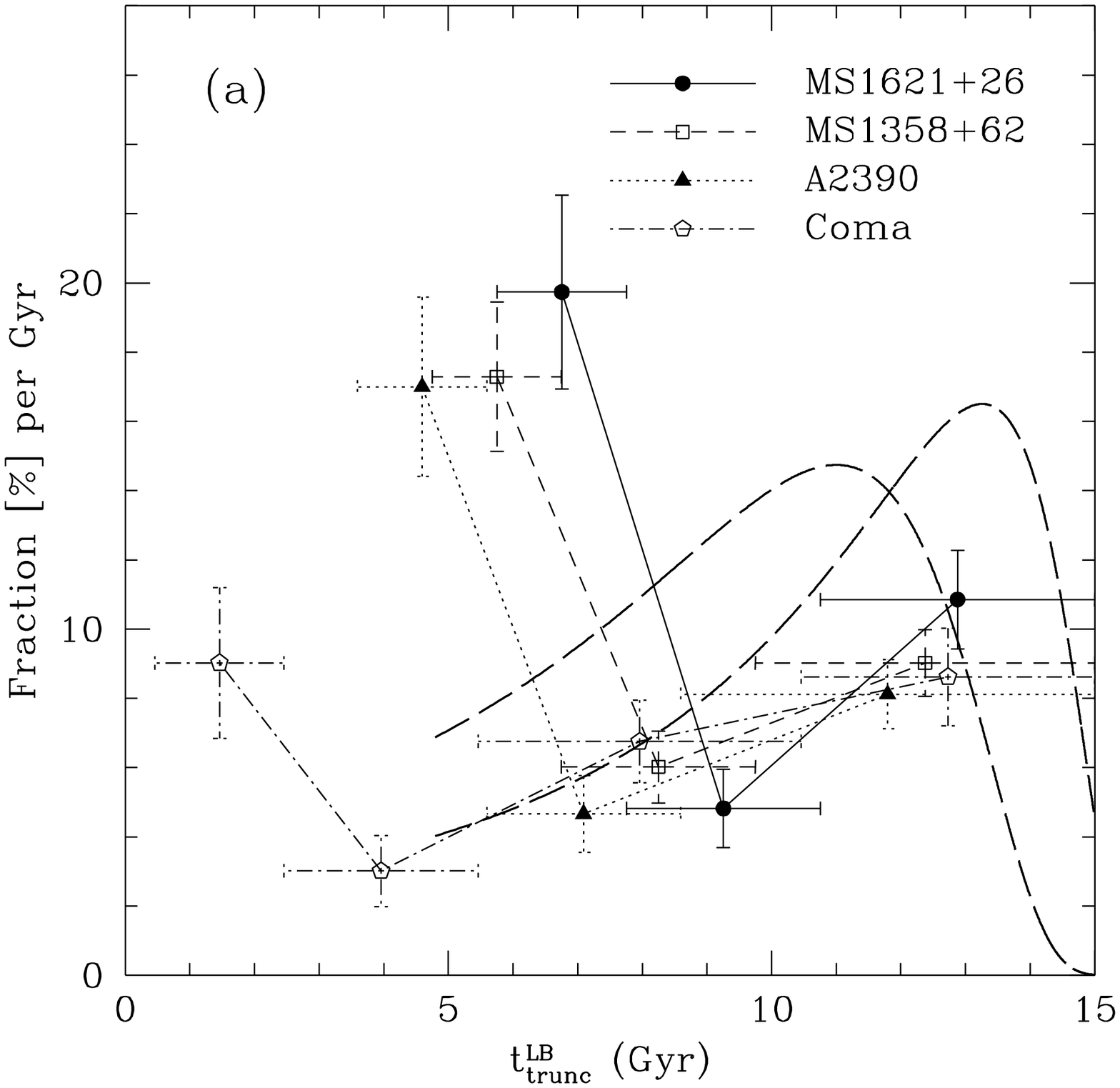}
  \epsfxsize 0.48\hsize
  \epsffile{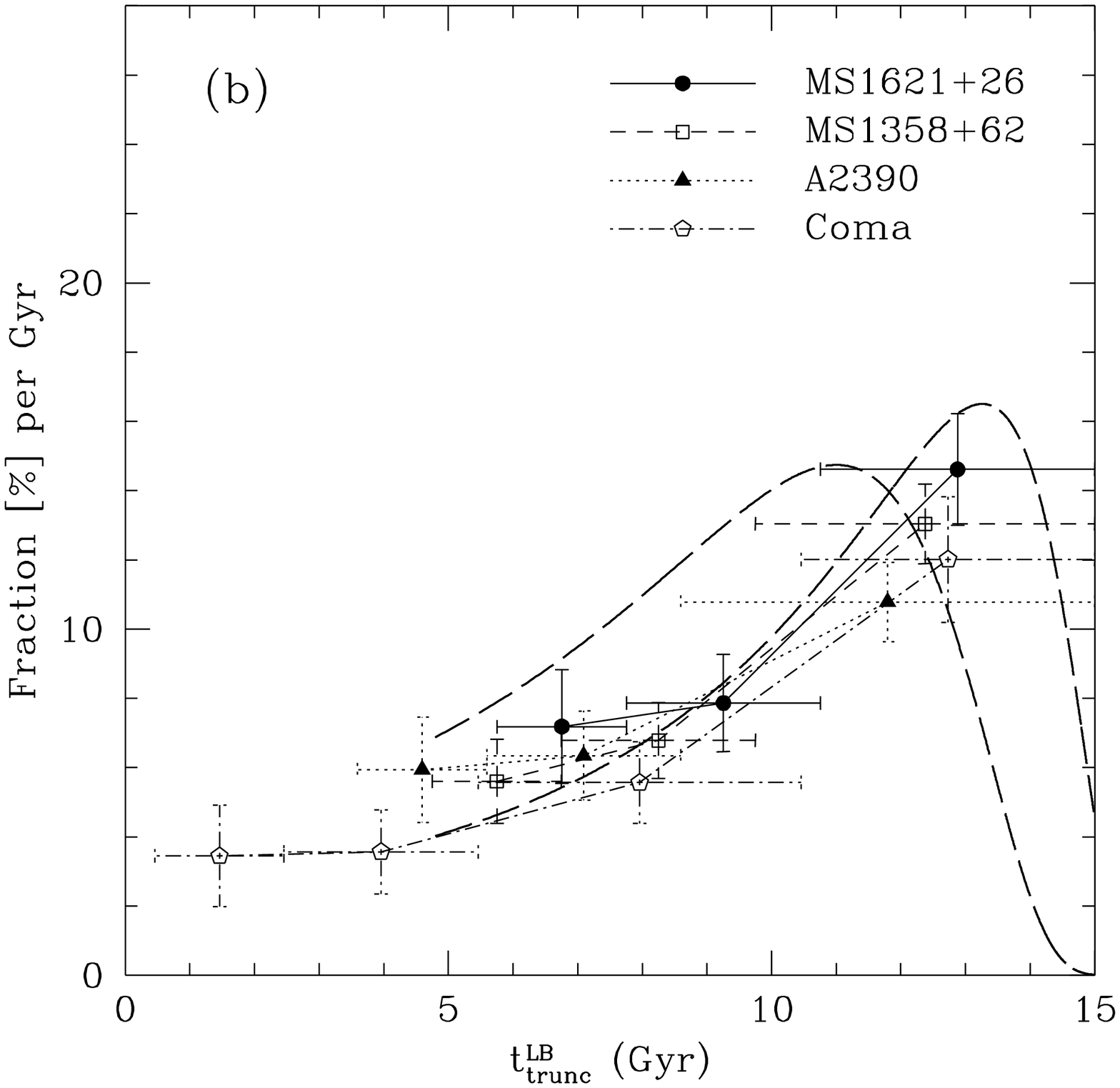}
\end{center}
\caption
{
The distribution of the star formation truncation time
for the galaxies with $M_V$$<$$-$19 at $z$=0.
The truncation epoch $t^{\rm LB}_{\rm trunc}$ is measured from the
{\it present-day}
rather than from the epoch at which the cluster is observed.
The vertical axis gives the
percentage of galaxies with this truncation time per Gyr.
The bin sizes are set so that colour differences can be resolved
within the photometric errors.
Two cases are considered regarding the star formation at the time of galaxy
accretion.
In Panel~(a), we assume
that half of the remaining gas in the disk is lost out of the system
and the other half is turned into stars in a burst at the galaxy accretion,
and no residual star formation is present afterwards.
In Panel~(b), we assume that no gas is lost from the disk
and only 1 per cent of disk gas is converted into stars by a burst.
The remaining gas is consumed exponentially with an e-folding time
of 1~Gyr. Thick dashed curves show the predicted galaxy accretion history of a 
$10^{15}$~M$_\odot$ cluster at $z$=0.33 for
a CDM-like power in an open universe ($\Omega_0$=0.2, $\sigma_8$=0.8,
$\Gamma$=0.1). The two curves in each plot are for threshold 
galaxy group mass of
$5\cdot10^{12}$~M$_\odot$ (right) and $5\cdot10^{13}$~M$_\odot$ (left),
respectively.
}
\label{fig:tinfall}
\end{figure*}

We assign star formation histories for individual cluster galaxies
statistically, based on the observed colour distribution, using the field 
colour distribution to define the initial galaxy colours.
The resulting distribution of the assigned truncation times can now be used
to infer the history of galaxy accretion from the field into the cluster.
Figure~\ref{fig:tinfall} shows the resulting distribution of $t_{\rm trunc}$
for galaxies brighter than $M_V$=$-$19 at $z$=0 for the three distant
clusters, plus the nearby Coma cluster. This magnitude cut ensures that 
the photometry is complete in all of the clusters.
The results of our Monte-Carlo
simulations have been combined into three bins (four bins for Coma).
Galaxies in which 
star formation has been recently truncated can be more accurately
age-dated than older systems: thus the age ranges of the bins increase to 
reflect the accuracy of the photometric data.
The two panels differ in the treatment of the truncation event as
described above.

As can be seen, the burst model (Model~(a)) produces a skewed
accretion pattern in which the cluster accretion rate is always highest at the
epoch of observations.
This is not consistent with the smooth decline in the accretion we had
expected, nor is it consistent between the different clusters.
It is unlikely that this peak is caused by an erratic accretion
history since we see the same pattern in all of the clusters.  
The fundamental problem with Model~(a) is that the blue galaxies evolve
toward the red sequence so rapidly that the model struggles to produce
the observed numbers of blue galaxies.
We note, however, in the current model with $e_{\rm burst}$=0.5,
the typical burst strength ($f_{\rm burst}$) at the time of truncation
is only a few per cent in mass for those galaxies just arrived
in the clusters. (Even with $e_{\rm burst}$=1, the burst strength is
typically 5 per cent). In order to avoid the peak in the accretion
rate at the observed epoch, we could raise the burst
strength so that the truncation epoch is pushed away into the past to get
the same colour. However, this becomes effective only when we consider
very strong bursts of at least 20 per cent in mass for {\it all} the galaxies.
Under the current gas infall model scheme (Appendix~B), an external
gas supply is needed to produce such a strong burst,
since the required amount of gas is not available for most of the galaxies.

The behaviour of Model~(a) can be improved if we assume that galaxies
can orbit in the cluster for up to 1 Gyr (the cluster crossing time)
before undergoing a star burst. This tends to smooth out the peak seen
in panel (a) because we observe some galaxies prior to the truncation event.
However, the peaks seen just before the observed epoch persist since
the bin size is already larger (2~Gyr) than the time delay.

In the second panel, we show the decaying model (Model~(b)). In this
model, the history of accretion is much more consistent between the
different clusters. In all four systems the accretion rate declines
monotonically with redshift. It is interesting to compare the evolution
of the accretion rate with that predicted on the basis of extended P-S
theory (Bower, 1991, Bond et al., 1991).
As an example, the predicted galaxy accretion history for a $z$=0.33 cluster 
of mass 10$^{15}$~M$_\odot$ is shown by the dashed lines in the figure.
This has been estimated from the extended P-S theory using a threshold
halo mass of 5$\cdot$10$^{12}$~M$_\odot$ and 5$\cdot$10$^{13}$~M$_\odot$
(for the right and left peaks respectively) at which the galaxy
is counted as having made the transition from the field to cluster
environment (Bower 1991; Kauffmann 1995). As can be seen the first of
these models fits the data remarkably well.  It is also notable
that the accretion history inferred from the Coma cluster matches smoothly
onto that inferred from the higher redshift systems. This comparison
gives us fundamental insight into the origin of the Butcher-Oemler
effect (ie., the differences between high and low redshift clusters). Both
the declining activity of field galaxies, and the declining rate
at which galaxies are accreted from the `field' environment play an
important role in producing the consistency (both between the accretion
histories of the clusters and between the reconstructed histories and the
theoretical model) that we see in this diagram.

\subsection{Comparison with Spectroscopic Studies}

\begin{table}
\caption{
The fraction (\%) of H$\delta$ strong galaxies and [\oii] emission line 
galaxies in the cluster cores. 
The MORPHS numbers and the CNOC numbers are taken from Poggianti et al. (1999) 
and Balogh et al. (1999), respectively.
The `corrected' CNOC numbers take into account the scatter due to 
measurment errors.
}
\label{tab:spectra}
\begin{tabular}{lccc}
\hline
obs/model & H$\delta$ ($>$3\AA) & H$\delta$ ($>$5\AA) & [\oii] ($>$5\AA) \\
\hline
MORPHS         & 21 $\pm$ 2  & 12 $\pm$ 1    & 30 $\pm$ 5 \\
CNOC           & 7.5 $\pm$ 1 & 4.4 $\pm$ 0.7 & 20 $\pm$ 5 \\
CNOC corrected & $<2.5$      & 1.5 $\pm$ 0.8 &  ---       \\
Model (a)      & 6           & 4             & 0$^\dag$   \\
Model (b)      & $<3$        & $<1$          & 13         \\
\hline
\end{tabular}\\
Note -- $^\dag$ If there is a time delay of 1~Gyr for the star burst
epoch with respect to the accretion epoch, this fraction increases to 12~\%..
\end{table}

So far, we have concentrated on reproducing a consistent accretion pattern
between the clusters based on the colour distribution of galaxies.
In this section, we compare our results obtained based on the photometric work
with the recent spectroscopic studies of galaxy evolution in clusters.
In order to estimate the fraction of the H$\delta$ strong (k+a/a+k) galaxies
in our models, 
we calculated the fraction of galaxies which have truncated their star
formation shortly before the observed epoch: ie.,
$t_{\rm obs}$$-$$t_{\rm trunc}$$<$1.5~Gyr (equivalent width
$W_0$(H$\delta$)$>$3~\AA) or
$t_{\rm obs}$$-$$t_{\rm trunc}$$<$1~Gyr ($W_0$(H$\delta$)$>$5~\AA),
and $f_{\rm gas}(t_{\rm trunc})$$>$0.1, where the calibration is 
based on Poggianti et al. (1999) and Balogh et al. (1999).
These two definitions of the H$\delta$ strong galaxies correspond to the
different definitions adopted by these two papers.
We also estimated the fraction of galaxies with [\oii] in emission, using
Kennicutt's (1992b) calibration:
\begin{equation}
{\rm SFR} [{\rm M}_{\odot} {\rm yr}^{-1}]>6.75 \times 10^{-12}
\frac{L_B}{L_{B,\odot}} W_0(\oii),
\end{equation}
where $L_B$ and $L_{B,\odot}$ are the blue luminosity of a galaxy and the
sun, respectively. 
The resulting fraction of  H$\delta$ strong galaxies and
emission galaxies  are summarised in 
Table~\ref{tab:spectra}, where they can be compared with the  
fractions observed by the MORPHS (Poggianti et al. 1999) and the CNOC 
(Balogh et al. 1999) collaborations.
For this comparison, we have transformed the numbers between the 3 and 
5~\AA\ criterion using the same factor given by Balogh et al. (1999).
A large difference in the fraction of H$\delta$ strong galaxies 
is evident between the MORPHS and the CNOC numbers, with the MORPHS values 
being higher by factor 3.
This difference is still an open question, although the possibilities are
extensively discussed in Balogh et al. (1999) including the intrinsic
difference between the optical (MORPHS) and the X-ray (CNOC) selection of
clusters.
Balogh et al. also presented the `corrected' fraction of
H$\delta$ strong galaxies, which has been corrected for the scattering effect
due to the measurement errors. This effect reduces the fraction by about
1/3.

Our Model~(b) gives numbers which are consistent with both
the CNOC corrected fraction of H$\delta$ strong galaxies and the 
fraction of [\oii] emission galaxies.
This suggests that Model~(b) is a good representation
of the CNOC results even from spectroscopic point of view.
Model~(b) fails to reproduce the fraction of H$\delta$ strong galaxies
found by MORPHS. 

In contrast, if we assume that the same correction for the scattering 
effect can be applied to the MORPHS numbers,
they are in good agreement with Model~(a). However, Model~(a) does not
produce any [\oii] emission galaxies because of the sharp truncation of star
formation, hence no on-going star formation at the observed epoch.
To reconcile this discrepancy, we allow for a time delay between the
accretion of the galaxy into the cluster and the star-burst/truncation
event. If the delay is 1~Gyr (the cluster dynamical time), 
the fraction of [\oii] emission galaxies could be up to 12 per cent: still
a factor of 2--3 lower than the observed fraction, but within the present
uncertainties.

\subsection{The History of Star Formation in Cluster Galaxies}

\begin{figure*}
\begin{center}
  \leavevmode
  \epsfxsize 0.48\hsize
  \epsffile{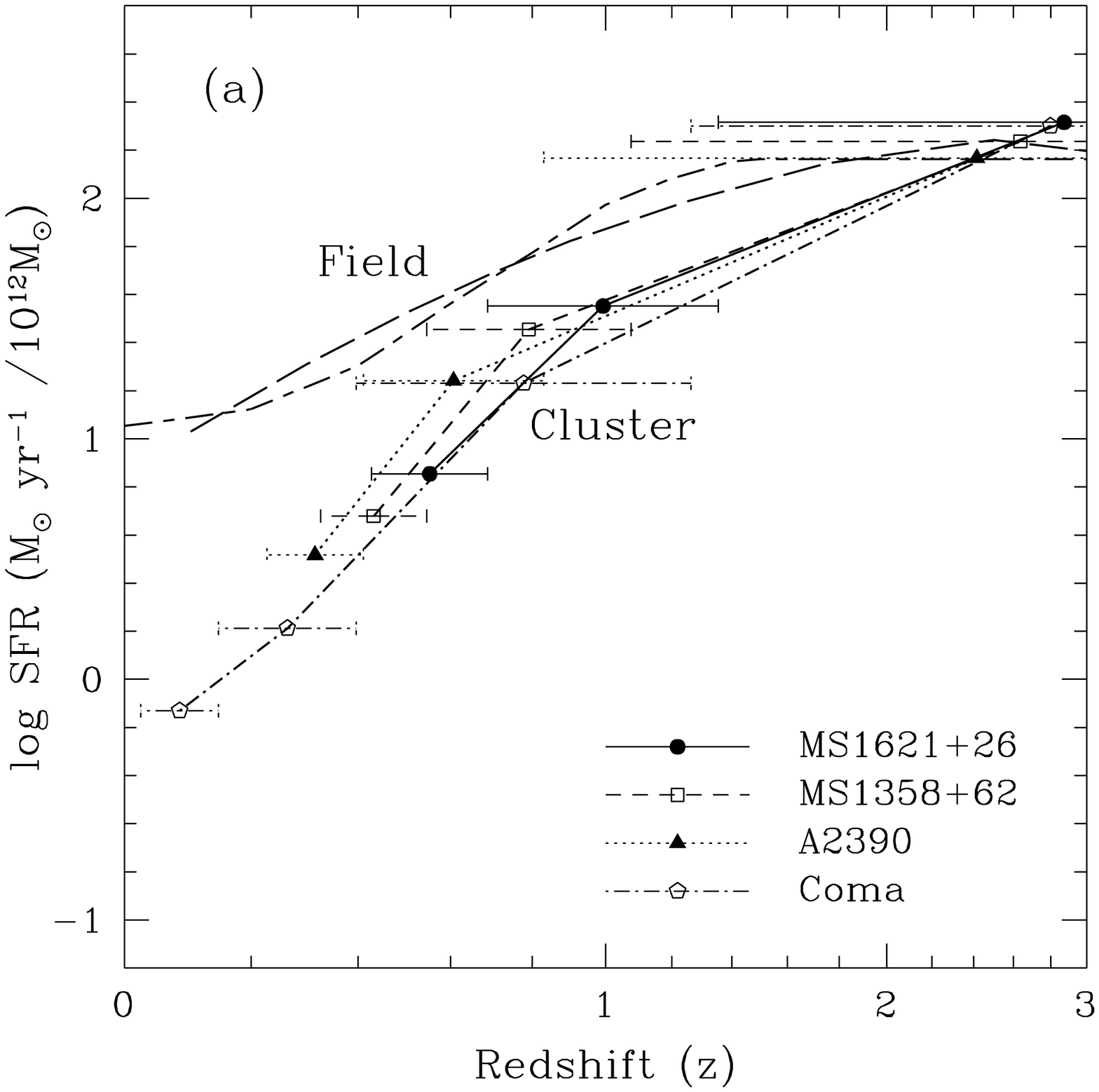}
  \epsfxsize 0.48\hsize
  \epsffile{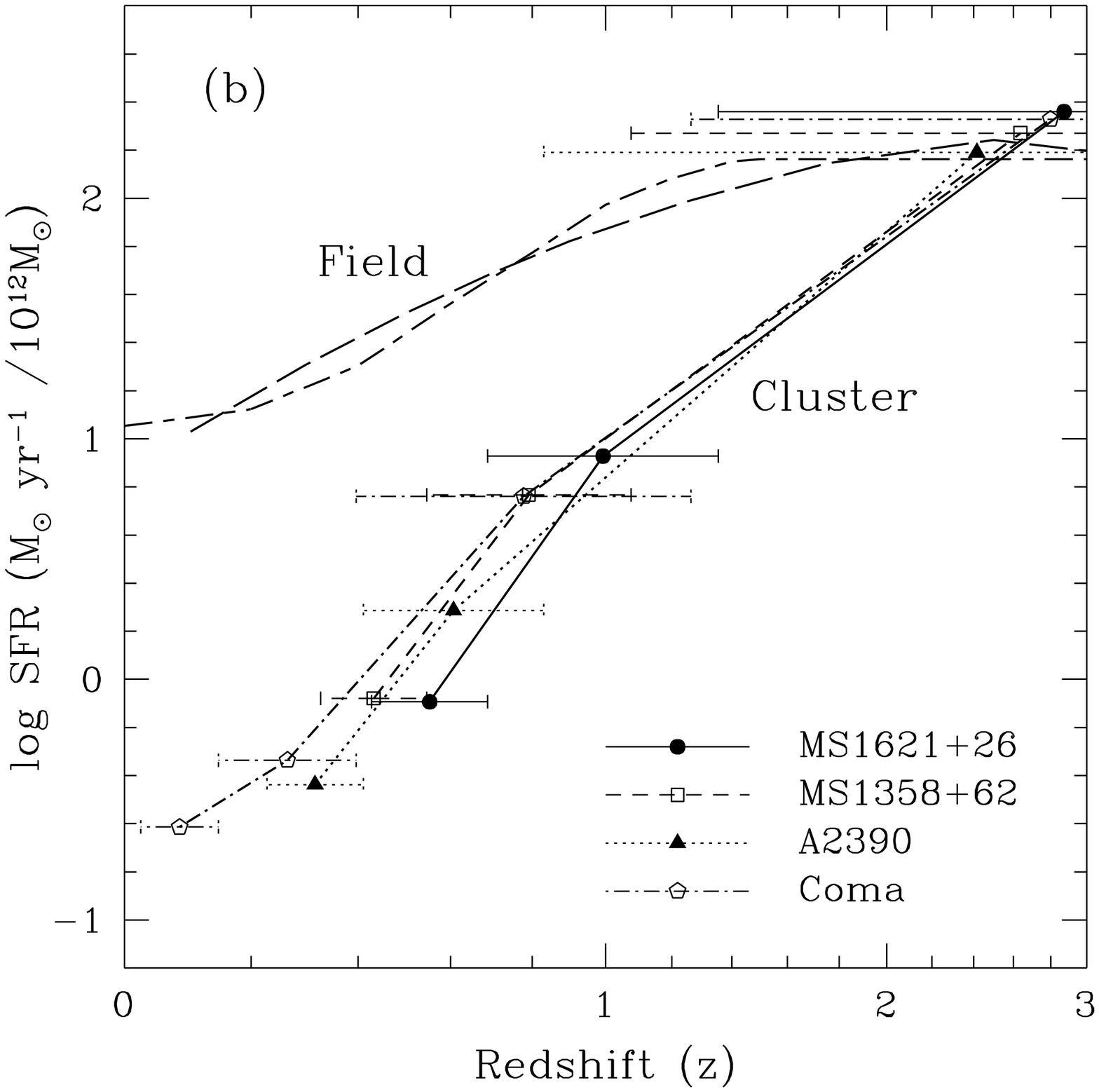}
\end{center}
\caption
{
The star formation rate as a function of redshift for galaxies in 
rich cluster cores. The star formation rates have been derived 
from  Model (a) (left panel) and Model (b) (right panel).
The star formation rate (SFR) is presented per galaxy mass of
10$^{12}$~M$_{\odot}$.
Four clusters are plotted individually by the curves with horizontal error
bars showing the bin
sizes; Coma (dot-dashed), A2390 (dotted), MS1358+62 (solid), and MS1621+26
(dashed).
The long dashed line above indicates the star formation history for the
CNOC field galaxies.
The long-short dashed line above show the cosmic star formation history (field)
taken from Madau et al. (1998).
}
\label{fig:sfh}
\end{figure*}

Finally, we use our approach to compare the global star formation
history in the cluster environment with that of the universe as a whole.
The global star formation history of the universe has received extensive 
attention since Madau  et al. (1996) combined data on the
local H$\alpha$ luminosity (Gallego et al. 1995), the UV luminosity
density from the CFRS
survey (Lilly et al. 1995) and the Lyman Break galaxies (Steidel et al. 1995).
This paper is a first attempt to show a similar plot for the galaxies in
the {\it cluster} environment.
Although we do not have direct observations at $z$$>$0.5, the data still
probe star formation histories 
out to $z$$\sim$1 due to reasonably good sensitivity of the rest
frame $U$$-$$V$ colour.
For each cluster, we sum up individual star formation histories over all the
cluster members (except the brightest cluster galaxy) and
normalised to a galaxy mass of 10$^{12}$~M$_{\odot}$.
In Fig~\ref{fig:sfh} we show the integrated star formation histories
for the galaxies in the rich cluster cores within $R_{30}$
for the four clusters (Coma, A2390, MS1358+62, and MS1621+26).
Two panels correspond to the two extreme models (a: burst model,
b: slow decay model) as in Fig.~\ref{fig:tinfall}.

The long-short dashed and the long dashed curves show the evolution of the
field galaxy star formation as suggested by Madau et al (1998) and as fitted
by our field galaxy model (Appendix B). We have assumed that the 
global star formation rate  is constant beyond $z$=1.5.

In the cluster cores, star formation at low redshifts declines much more
rapidly than in the field, with galaxy formation skewed to higher redshifts.
At $z=0.5$, the typical star formation rate per galaxy mass is a
factor 3--10 smaller in cluster cores, depending on the model chosen.
This is in qualitative agreement with Balogh et al. (1998) and Couch et al.
(2000) who find a strong decline in the star formation rate from the field 
to the cluster environment at intermediate redshifts.
Although we prefer Model~(b), which gives the most consistent picture of the
assembling history of cluster galaxies,
we should note that the exact locus of cluster star formation
history on this plot depends on the treatment of the truncation, especially
on the amount of gas that is lost from the system when it is accreted into 
the cluster (ie., $e_{\rm loss}$).
This uncertainty dominates the highest redshift bin since we have only a
model dependent constraint (from the total number of galaxies now lying on the
red sequence) on the accretion rate long before the cluster is observed.
The important next step is to directly map the accretion history at higher 
redshifts by extending the method to more distant clusters.
In addition, better understanding of the way in which star formation is 
truncated in the clusters and within individual galaxies eg., from H$\alpha$ 
star formation surveys (eg., Balogh \& Morris 2000;
Couch et al. 2000) will better constrain the modelling that is required 
to determine the cluster infall rate.

\section{Conclusions}

We have created the field corrected CM diagrams for the 7 CNOC distant
clusters (0.23$\lsim$$z$$<$0.43) in order to study their galaxy populations
in the same manner as in BO84. We make a comparison with
the redshift using data for the Coma cluster from 
Terlevich et al. (2000). The form for the redshift evolution
of the blue galaxy fraction agrees well with the trend seen in BO84.
In addition, the blue galaxy fraction is generally a strong 
increasing function of distance from the cluster centre, and depends
only weakly on magnitude.

We have then applied these data-sets to build up a sequence of snap-shots
of cluster evolution. We follow the evolution of the whole population 
of cluster galaxies in these CM diagrams accounting for the flow
of galaxies across the diagram as they are accreted by the cluster.

\begin{itemize}

\item 
Firstly, we have shown that the blue galaxies are incorporated into the 
present-day tight CM relation as they fade and become redder rapidly after the
truncation of their star formation, giving the possible link to
the faint ($>$$M_*$+1) S0 galaxies in present-day clusters.
The brighter S0 galaxies, however, must have ceased star formation
earlier than $z\sim0.4$. In clusters, as in the universe as a whole,
fainter galaxies have a more extended star formation history than their
bright counterparts. Nevertheless,
this effect does not lead to a sizable age variation along the
CM relation at the present-day.

\item Secondly, we have used the distribution of galaxies in the colour
space to infer the evolution of the rate at which galaxies have been 
accreted by the cluster. We interpret the observed distribution in
colour by following the declining star formation rate in the field
and statistically assigning a star formation truncation epoch.
If star formation is truncated abruptly, we find that too many recent
accretion events are required relative to the past rate.
In order to be consistent with a smoothly declining accretion rate,
such as suggested by the extended P-S theory,
we need to allow for some level of residual star formation after galaxy 
accretion. This suggests that ram-pressure and/or mergers/harassment 
cannot be so effective as to remove all the remaining gas from the system
in a single orbital time-scale. The Butcher-Oemler effect is thus shown to 
result from a combination of the three effects,
namely, declining accretion rate, declining field star formation rate,
and the truncation of star formation.

\item Finally,
we have presented the global star formation history for the galaxies
in rich cluster cores for the first time.
It is shown that the star formation in cluster galaxies is
much less than that in the field galaxies below $z$=1. However, the factor 
by which star formation is suppressed depends on the effectiveness
of the cluster environment in suppressing star formation.

\end{itemize}

It is important to ask how robust these conclusions are to the assumptions 
that must be made in order to model the galaxy properties. When
propagating the observed properties of galaxies forward in time, the
past star formation rate has only a relatively weak effect (as shown by
Fig.~\ref{fig:vector}) and thus the results derived from this analysis
are robust. Determining the distribution of truncation times, and hence
comparing the past and present accretion rates in clusters is
significantly more model dependent. Firstly, the ratio of the present to 
past star formation rate must be estimated for the field galaxy
population. This must be parameterised by a simple model and matched to
the observed colour distribution of field galaxies. The properties of
the accreting galaxies can, however, be constrained by a wide range of
observations in addition to the observed field galaxy colours. The need to
match the cosmic star formation history suggested by Madau et al. (1998) 
has motivated our
choice of the gas infall model. Secondly, we must parameterise the effect of
the cluster environment on the galaxies that are accreted. This is the
more critical component of the model, and we illustrate this by
considering two scenarios. If star formation is cut-off abruptly by a 
star burst and gas stripping upon accretion into the cluster,
the model struggles to explain
the high fraction of blue objects seen in the distant clusters. In
contrast an extended period of star formation decay naturally reproduces
a smooth decay in the accretion rate that is similar in all of the
clusters. More complex scenarios are easily possible: for example
galaxies might either undergo a star burst or might decay slowly,
depending on their orbit within the cluster. 
An important avenue for future work is to integrate spectral
star formation indicators (such as the [\oii] or H$\alpha$ line strength), 
or even radio continuum flux (Smail et al. 1999),
and H$\delta$ absorption
line strength into this analysis so that the degeneracy between colour
and past/present star formation rate can be broken. This will improve our
understanding of the truncation process.

So far we have not considered the effects of dust obscuration. This is
a potentially important complication that is extremely difficult to
model in a straightforward manner. Adding a uniform dust extinction that 
depends on the present star formation rate reddens the observed colours, 
causing us to underestimate star formation rates. However, both the field
and some of the cluster galaxies will be affected so that the reddening
cancels out to a first approximation. On the other hand,
if localised dust extinction accompanies a star-burst
on entry to the cluster environment as suggested by Poggianti et al. (1999)
and Smail et al. (1999), we would recover an uneven accretion 
history (similar to the behaviour shown by Model~(a)) since 
the `intrinsic' (ie., dust corrected) colour distribution would include 
many more `blue' galaxies.

Clearly the potential of these techniques are best exploited by
observing more and more distant clusters. 
To resolve the star formation history in clusters at higher redshifts
($z$$>$1), we need a good homogeneous sample of clusters at higher
redshifts ($z$$>$0.5--1.0).
The {\it Rosat} deep X-ray cluster survey (Rosati et al. 1998)
and the coming new surveys by
{\it Chandra} and {\it Newton} with high sensitivity cameras will provide
an ideal sample of high redshift clusters for this purpose.

The analysis presented in this paper is limited by the accuracy of the 
the distant cluster photometry, as well as by the number of clusters 
in our sample.
A larger sample of clusters at each redshift would allow
projection effects to be averaged over and the variation in evolution 
history to be investigated.
Wider spatial coverage is also important to more accurately subtract field
population and also to investigate the environmental effect beyond the cluster
core out to the virial radius. This is essential if we are to improve
our knowledge of the star formation truncation mechanism itself. At present,
our approach is also compromised because we do not allow for the
dynamical evolution of the cluster populations. Because of this, we cannot
compare the properties of galaxies at different radii within the clusters.
Recent rapid progress in the high resolution N-body simulation of clusters
(eg., Moore et al. 1999b; Balogh et al. 2000; Diaferio et al. 2000)
will allow this aspect to be taken into account.

\section*{Acknowledgements}

We are indebted to B. Abraham, H. Yee, and other CNOC group members for
providing us their photometry catalogs and A. Terlevich for the Coma data
which form the base of this work.
We also thank I. Smail, M. Balogh, W. Couch, B. Poggianti and E. Bell for
valuable discussions.
We are grateful to the anonymous referee who gave us many useful comments
which improved this paper.
T.K. thanks JSPS Research Fellowships for Young Scientists
for financial support, and Univ. of Durham for kind hospitality during
his stay in Durham.
This project has made extensive use of Starlink 
computing facilities in Durham.

\appendix

\section{Statistical Field Galaxy Subtraction}

This appendix describes our method for subtracting the
foreground/background populations from the cluster fields.
The sample of field galaxies is taken from the edge of the cluster fields
of three CNOC clusters with wide imaging fields; ie.,
\begin{center}
\begin{tabular}{cl}
$11<\Delta {\rm RA}<21$ & for A2390, \\
$8<\Delta {\rm DEC}<11.5$ & for MS1358+62, \\
and $-12<\Delta{\rm DEC}<-8$ & for MS1621+26, \\
\end{tabular}
\end{center}
where $\Delta{\rm RA}$ and $\Delta {\rm DEC}$ are the distances from
the cluster centre in the unit of arcmin, and these correspond to $\gsim$3~Mpc
which is comparable to or slightly smaller than the virial radius
(Carlberg et al. 1996). 
We omit the MS1512+36 cluster because of probable contamination by a
foreground system. We can estimate the contamination of the `field' 
sample either by extrapolating the surface density profile of clusters \
or directly from the CNOC spectroscopic data. Both methods suggest that only 
10--15 per cent of the galaxies are expected to be
associated with the outer regions of the clusters.
After rejecting a few spectroscopically confirmed cluster members from this
field area, the total number of the field galaxies is 871 down to $r$=23.5.
Divided by the total area of 137.5 arcmin$^2$,
it gives the field number density of 6.34$\pm$0.21 galaxies per arcmin$^2$.
This can be compared to the deep field galaxy counts in Gunn's system
(Brainerd et al. 1995)
which has 90.1 arcmin$^2$ field of view and 7.01$\pm$0.28 galaxies per
arcmin$^2$ with the same magnitude cut.
Furthermore, both the magnitude and colour distribution of galaxies are similar
between the two fields.
Therefore our field sample can be regarded as a good representation of the
field population.
Even if we used Brainerd et al.'s (1995) field population, the results
presented in this paper would hardly change.
The CNOC field galaxies defined in this way are plotted on the CM diagram
in the bottom right panel of Fig.~\ref{fig:cm2}.

We build the two dimensional distribution histogram of these field galaxies
with bins of width 0.3 in colour and 0.5 in magnitude.
A similar histogram is also created for each cluster field within $R_{30}$.
This obviously includes the field galaxies that are to be subtracted.
The field histogram is normalised for each cluster to match the area within
$R_{30}$.
The number of spectroscopically confirmed non-members is then subtracted
from these two histograms in each bin. The resulting number can be negative.
We now count the number of galaxies in each bin ($i$,$j$) for the field
($N_{ij}^{\rm field}$) and the cluster plus field ($N_{ij}^{\rm cluster+field}$).
The former can be larger than the latter for low density bins due to
low number statistics.
If this happens, we re-distribute the excess number of field galaxies
to the neighbouring bins (8, 5, or 3 bins depending on the
location of the bin under concern) with equal weight until
\begin{equation}
N_{ij}^{\rm field} < N_{ij}^{\rm cluster+field}
\end{equation}
is satisfied in all bins. 
We then define a probability, $P$, for each galaxy in the cluster field
that it is a field galaxy:
\begin{equation}
P=\frac{N_{ij}^{\rm field}}{N_{ij}^{\rm cluster+field}},
\end{equation}
where the numbers are taken from the bin that the galaxy belongs to.
We now apply Monte-Carlo simulations to statistically determine the field
galaxies.
We randomly pick out a membership-unknown galaxy from the entire CM diagram
and refer to its probability $P$ to determine whether it should be regarded
as a field galaxy.
We repeat this process for all galaxies in the cluster field and
rerun the Monte-Carlo simulation to average
over the 100 realisations for each cluster.

\section{The Infall Model}

We use the gas infall model prescription for star formation
in galaxies under the instantaneous recycling approximation (Tinsley 1980).
We do not take into account the chemical evolution in the galaxy system
in this paper, and the metallicity is fixed at solar value.
This treatment is supported by BKT98, where they found the colour changing
rate is only a weak function of metallicity.

The galactic gas is supplied from a surrounding gas reservoir (halo)
at the rate:
\begin{equation}
\xi_{\rm in}(t)=\frac{1}{\tau_{\rm in}}\exp\left(-\frac{t}{\tau_{\rm in}}\right).
\end{equation}
The total mass of the gas in the reservoir is normalised to unity.
The zero-point of the time $t$ is set to the look back time of 15~Gyr which
corresponds to the redshift of 5.4 in our cosmology.

A Schmidt law (Schmidt 1959) with a power unity is adopted,
where star formation rate is proportional to the gas density, hence mass, as
\begin{equation}
\psi(t)=\frac{1}{\tau_*}M_{\rm gas}(t),
\end{equation}
where $M_{\rm gas}$ indicates the mass of the gas that has infallen and has
not yet been turned into stars. The initial gas mass, $M_{\rm gas}(0)$, 
is set to zero.
Note that this formulation gives an exponentially decaying star formation rate
with an effective time scale $\tau_*^{\rm eff}=\tau_*/\alpha$ in the case of
the simple models,
where $\alpha$ is the so-called locked-up mass fraction defined by Tinsley
(1980). The Salpeter mass function (Salpeter 1955)
that we adopted in this paper with lower mass cut-off 0.1~M$_{\odot}$
and upper mass cut-off 60~M$_{\odot}$, gives $\alpha=0.72$.

In order that the gas fraction is stable, we adopt
\begin{equation}
\tau_*^{\rm eff}=\tau_{\rm in} \equiv \tau.
\end{equation}
This assumption leads to the simple analytic solutions for the star formation
in a galaxy:

\begin{equation}
\psi(t)=\frac{1}{\alpha}\frac{t}{\tau^2}\exp\left(-\frac{t}{\tau}\right),
\end{equation}
which has a peak at $t=\tau$.

The gas mass and stellar mass are given by:
\begin{equation}
M_{\rm gas}(t)=\frac{t}{\tau}\exp\left(-\frac{t}{\tau}\right)
\end{equation}
and
\begin{equation}
M_*(t)=1-\left(1+\frac{t}{\tau}\right)\exp\left(-\frac{t}{\tau}\right),
\end{equation}
respectively.
We also define the gas mass fraction as:
\begin{equation}
f_{\rm gas}(t)=\frac{M_{\rm gas}(t)}{M_{\rm gas}(t)+M_*(t)}.
\end{equation}

Equations from (B1) to (B6), except (B3), are scaled for each galaxy to
match its absolute magnitude.

\section{Truncation and Star Burst}

When a galaxy is accreted from the field environment by the cluster
($t_{\rm acc}$),
all the gas in the halo is assumed to be stripped off due to tidal
stripping and the ram-pressure of the intra-cluster medium.
This prevents further gas being supplied from the halo gas reservoir.
A substantial fraction of the gas in a galaxy disk (ie., the material which
has been accreted from the halo) 
may also be stripped away by the ram-pressure
as numerical simulations suggest (Abadi et al. 1999;
Quilis et al. 2000).
The mass of gas that is lost from the system is described using
the remaining gas mass and the gas loss efficiency parameter
$e_{\rm loss}$ as
\begin{equation}
M_{\rm loss}=M_{\rm gas}(t_{\rm acc}) \times e_{\rm loss}.
\end{equation}

We allow a star burst to accompany the gas stripping process.
Gas mass that is transformed into stars in the burst is
described by a burst efficiency parameter $e_{\rm burst}$ and is given by
\begin{equation}
M_{\rm burst}=M_{\rm gas}(t_{\rm acc}) \times e_{\rm burst}.
\end{equation}
The burst strength is represented by the mass fraction of
the burst population to the total stellar mass:
\begin{equation}
f_{\rm burst}=\frac{M_{\rm burst}}{M_*(t_{\rm acc})+M_{\rm burst}}.
\end{equation}

After these processes, star formation in the galaxy is rapidly truncated
at time $t_{\rm trunc}$.
We assume that the star formation truncation time $t_{\rm trunc}$
is approximately equal to the galaxy accretion time $t_{\rm acc}$.
After $t_{\rm trunc}$, the galaxy quickly consumes the
remaining gas in the disk following a Schmidt law:
\begin{equation}
\psi(t)=\frac{1-e_{\rm loss}-e_{\rm burst}}{\alpha\tau_{\rm trunc}}
M_{\rm gas}(t_{\rm trunc})\exp\left(-\frac{t-t_{\rm trunc}}{\tau_{\rm trunc}}\right)
\end{equation}
where $\tau_{\rm trunc}$ is a time scale for the truncation of star formation
and is fixed to 1~Gyr.  In principle the model could be adjusted to use a 
range of decay timescales. We adopt 1~Gyr since it is midway between the
average timescale for field galaxies and abrupt truncation.

We consider two specific models in this paper that are representative of
the extreme possibilities:
\begin{center}
\begin{tabular}{ccc}
Model & $e_{\rm loss}$ & $e_{\rm burst}$ \\
(a) & 0.5 & 0.5  \\
(b) & 0.0 & 0.01
\end{tabular}
\end{center}
Model (a) is motivated by Quilis et al. (2000) and reflects what would
happen if the diffuse hydrogen were removed from the galaxy by ram-pressure
stripping while the remaining molecular gas (typically 50~\% of the total
gas mass) is consumed in a pressure induced star burst. Model (b) mimics
the decline in star formation rate expected if the action of the cluster
is simply to remove the outer gas reservoir.

\section{Assigning the star formation time scale and the truncation epoch}

\begin{figure}
\begin{center}
  \leavevmode
  \epsfxsize 1.0\hsize
  \epsffile{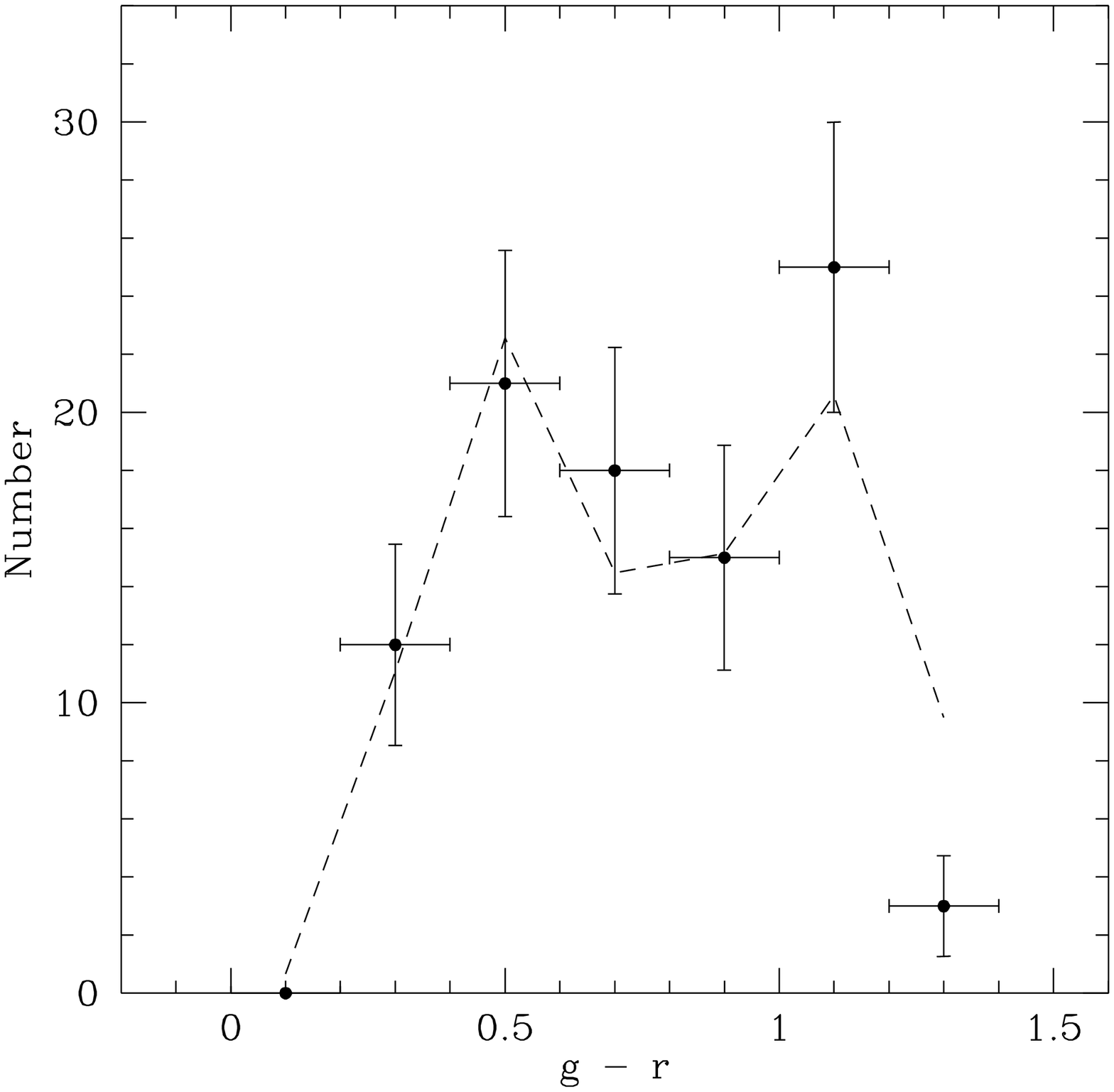}
\end{center}
\caption
{
Distribution of $g$$-$$r$ colour for the spectroscopically confirmed field
galaxies which are located between 0.3$<$$z$$<$0.35. These galaxies are taken
from all of our 7 CNOC cluster fields and the total number amounts to 94.
The dashed line corresponds to the Gaussian distribution of $\tau$ for the
field galaxies in equation~(D1).
}
\label{fig:field_gr}
\end{figure}

This section is devoted to explaining how we assign the star
formation/gas infall time scale ($\tau$) and the truncation epoch
($t_{\rm trunc}$) for each cluster galaxy.

Since we consider that the cluster galaxies is accreted from the field, the 
distribution of $\tau$
for the cluster galaxies is taken to be the same as for the field population.
This is directly estimated from the colour distribution of the real field
galaxies through our treatment of star formation described in Appendix~B.
Here we use 94 spectroscopically confirmed field galaxies between
0.3$<$$z$$<$0.35 from the entire observed fields of the seven CNOC clusters.
The estimated distribution of $\tau$ for the field galaxies can be
expressed by a Gaussian:
\begin{equation}
\Phi(\tau) \propto
\exp\left\{-\frac{1}{2}\left(\frac{\log\tau-\log\tau_0}{\sigma}\right)^2\right\},
\end{equation}
where $\log\tau_0$=0.3, and $\sigma$=0.2.
This distribution not only gives a reasonable match to the colour
distribution of the CNOC field galaxies (Fig.~\ref{fig:field_gr}),
but also reproduces the so-called Madau curve (Madau et al. 1998);
ie. the global star formation rate for field galaxies as a function of
redshift at the same time (see the long dashed curve against the long-short
dashed curve in Fig.~\ref{fig:sfh}).

When the galaxies fall into clusters, star formation ceases
with either gas loss and/or a star burst as described in Appendix~C.
Now, in order to reproduce a certain colour of a given cluster galaxy,
various combinations of $\tau$ and $t_{\rm trunc}$ are possible for a given
set of $e_{\rm loss}$ and $e_{\rm burst}$,
although $\tau$ should be larger than a certain minimum value
$\tau_{\rm min}$ which corresponds to $t_{\rm trunc}=0$.

For each possible combination of ($\tau$,$t_{\rm trunc}$),
the colour changing rate $R$ at the cluster redshift is calculated:
\begin{equation}
R(\tau,t_{\rm trunc}) \equiv \left.\frac{d(g-r)}{dt}\right|_{z=z_{\rm cluster}}.
\end{equation}
Note the exception that the colour used is $U$$-$$V$ for Coma.
The probability of assigning $\tau$, $q(\tau)$, is inversely proportional
to this rate since a galaxy with higher colour changing rate
is rarer to find:
\begin{equation}
q(\tau)=\left\{
\begin{array}{cc}
c/R(\tau,t_{\rm trunc}) & \tau \ge \tau_{\rm min}\\
0 & \tau < \tau_{\rm min}
\end{array}
\right.
\end{equation}
where the normalisation $c$ is determined by
\begin{equation}
\int q(\tau) d\log\tau= 1.
\end{equation}

The distribution of $\tau$ is thus skewed towards larger values due to the
minimum cut $\tau_{\rm min}$ and the $q(\tau)$ term.
To get the same distribution of $\tau$ in Equation~(D1) for the cluster
galaxies, the probability of assigning a given $\tau$ should be corrected
for these effects.
The correction function $Q(\tau)$ is the sum of the individual $q(\tau)$
term over all the cluster members:
\begin{equation}
Q(\tau)=\Sigma q_i(\tau).
\end{equation}

Finally, we assign $\tau$ randomly for each cluster galaxy using the
following distribution function $\Phi'(\tau)$:
\begin{equation}
\Phi'(\tau)=\Phi(\tau)/Q(\tau) \times q(\tau).
\end{equation}
The truncation time, $t_{\rm trunc}$, is then determined by requiring
the galaxy to have its observed colour at the cluster look-back time.

\end{document}